\documentclass{article}
\usepackage[letterpaper,hmargin=1in,vmargin=1.25in]{geometry}

\usepackage{cancel}
\usepackage{indentfirst}
\usepackage{amsfonts,bm}
\usepackage{fancyhdr}                               
\usepackage{amsmath,amscd,amssymb,latexsym}     
\usepackage{mathrsfs,mathcomp}
\usepackage{stmaryrd}
\usepackage{amsthm}
\usepackage[all,cmtip]{xy}  
\usepackage{titlesec}                               
\usepackage{caption}                              
\usepackage{cite}
\usepackage{epsfig,epic}

\usepackage{graphicx,color,xcolor,subfigure}
\usepackage{algorithmic}
\usepackage{algorithm}

\usepackage{tipa}
\usepackage{textcomp}

\newtheorem{theorem}{Theorem}
\newtheorem{lemma}{Lemma}   
{\theoremstyle{definition}

}

\makeatletter
\newcommand{\Rmnum}[1]{\expandafter\@slowromancap\romannumeral #1@} 
\makeatother
\newcommand{\set}[1]{\left\{#1\right\}} 
\newcommand{\cset}[1]{\left|#1\right|} 

\newcommand{\pf}{\begin{proof}}  
\newcommand{\fp}{\end{proof}} 

\newcommand{\Int}{\mathbb{Z}} 
\newcommand{\GF}[1]{\mathbb{F}_{#1}} 

\newcommand{\myNOT}{\neg} 
\newcommand{\myOR}{\vee} 
\newcommand{\myAND}{\wedge} 
\newcommand{\myXOR}{\oplus} 
\newcommand{\BnrSet}{\{0,1\}}

\newcommand{\crct}[1]{#1} 

\newcommand{\sizeof}[1]{\mathbf{SIZE}\left(#1\right)} 
\newcommand{\lengthof}{\per}

\newcommand{\cascade}{*} 

\newcommand{\state}[1]{
\mathbf{#1}} 
\newcommand{\vecone}[1]{\mathbf{1}^{#1}} 
\newcommand{\vecnull}[1]{\mathbf{0}^{#1}} 
\newcommand{\veconenull}[1]{\bm{\iota}^{#1}} 

\newcommand{\lbit}[2]{\lfloor{#2}\rfloor_{#1}} 
\newcommand{\hbit}[2]{\lceil{#2}\rceil_{#1}} 
\newcommand{\veccomplement}[1]{\overline{#1}} %
\newcommand{\vecconjugate}[1]{\widehat{#1}} %

\newcommand{\cycle}[1]{{\bm{#1}}} 
\newcommand{\CycleSet}[1]{\mathfrak{#1}} 
\newcommand{\CycleStr}[1]{\mathbf{CycStr}\left(#1\right)} 
\newcommand{\seqcyc}[1]{\theta\left(#1\right)} 
\newcommand{\cycsecset}[2]{S_{#1}\left({#2}\right)} 

\newcommand{\cycchar}[1]{\chi\left({#1}\right)} %
\newcommand{\cycctrl}[1]{\pi\left({#1}\right)} %
\newcommand{\cycextr}[1]{\lambda\left(#1\right)} %

\newcommand{\SeqSet}[1]{G\left(#1\right)} 
\newcommand{\seq}[1]
{#1} 
\newcommand{\per}[1]{\mathrm{per}\left(#1\right)} 

\newcommand{\NP}{$\mathbf{NP}$} 

\newcommand{\len}{n} %
\newcommand{\IsEqual}[1]{\stackrel{\text{\tiny #1}}{=}} %

\newcommand{\veccyc}[1]{\xi\left(#1\right)} 
\newcommand{\cycvec}[1]{\rho\left(#1\right)} 

\newcommand{\relint}[1]{\langle#1\rangle} 

\newsavebox{\register}
\savebox{\register}(0,0)[bl]{
\put(9,5){\vector(0,1){3.5}}
\put(5,0){\framebox(8,5){}}
\put(0,2.5){\vector(1,0){4.9}}
}

\newsavebox{\myLFSRpj}
\savebox{\myLFSRpj}(0,0){ 
\put(0,2.5){\vector(1,0){4.9}}
\put(5,0){\framebox(8,5){}}
\put(13,2.5){\vector(1,0){4.9}}
\put(20.5,2.5){\makebox(0,0){$\cdots$}}
\put(23,0){\usebox{\register}}
\put(32,10){\circle{3}}
\put(32,10){\makebox(0,0){$\myXOR$}}
\put(36,2.5){\vector(1,0){4.9}}
\put(43.5,2.5){\makebox(0,0){$\cdots$}}
\put(46,2.5){\vector(1,0){4.9}}
\put(51,0){\framebox(8,5){}}
\put(55,5){\line(0,1){5}}
\put(55,10){\vector(-1,0){21.5}}
\put(30.5,10){\line(-1,0){30.5}}
\put(0,10){\line(0,-1){7.5}}
\put(59,2.5){\vector(1,0){4.9}}
\put(65.5,2.5){\circle{3}}
\put(65.5,2.5){\makebox(0,0){$\myXOR$}}
\put(67,2.5){\vector(1,0){4.9}}
\put(72,0){\framebox(8,5){}}
\put(76,5){\line(0,1){5}}
\put(76,10){\line(-1,0){10.5}}
\put(65.5,10){\vector(0,-1){6}}
\put(80,2.5){\vector(1,0){4.9}}
}

\newsavebox{\myLFSRjp}
\savebox{\myLFSRjp}(0,0){ 
\put(0,2.5){\vector(1,0){4.9}}
\put(5,0){\framebox(8,5){}}
\put(9,5){\line(0,1){5}}
\put(9,10){\line(-1,0){9}}
\put(0,10){\line(0,-1){7.5}}
\put(13,2.5){\vector(1,0){4.9}}
\put(19.5,2.5){\circle{3}}
\put(19.5,2.5){\makebox(0,0){$\myXOR$}}
\put(21,2.5){\vector(1,0){4.9}}
\put(26,0){\framebox(8,5){}}
\put(34,2.5){\vector(1,0){4.9}}
\put(41.5,2.5){\makebox(0,0){$\cdots$}}
\put(44,0){\usebox{\register}}
\put(53,10){\circle{3}}
\put(53,10){\makebox(0,0){$\myXOR$}}
\put(57,2.5){\vector(1,0){4.9}}
\put(64.5,2.5){\makebox(0,0){$\cdots$}}
\put(67,2.5){\vector(1,0){4.9}}
\put(72,0){\framebox(8,5){}}
\put(76,5){\line(0,1){5}}
\put(76,10){\vector(-1,0){21.5}}
\put(51.5,10){\line(-1,0){32}}
\put(19.5,10){\line(0,-1){6}}
\put(80,2.5){\vector(1,0){4.9}}
}

\newsavebox{\myFSR}
\savebox{\myFSR}(0,0){
\put(0,0){\usebox{\register}}
\put(13,0){\usebox{\register}}
\put(26,2.5){\vector(1,0){4.9}}
\put(35,2.5){\makebox(0,0){$\cdots$}}
\put(39,0){\usebox{\register}}
\put(52,0){\usebox{\register}}
\put(65,2.5){\vector(1,0){4.9}}
\put(35,11){\oval(60,5)}
\put(0,11){\line(1,0){5}}
\put(0,11){\line(0,-1){8.5}}
}
\newsavebox{\myCPCLM} 
\savebox{\myCPCLM}(0,0){
\put(16,1.5){\oval(4,3)\makebox(0,0){$L$}}
\put(16,6){\vector(0,-1){3}}
}

\newsavebox{\myCPcircuit} 
\savebox{\myCPcircuit}(0,0){
\put(16,25.5){\makebox(0,0){\small input $\state{x}$ }}
\put(16,23.5){\circle*{0.55}}
\put(16,23.5){\vector(0,-1){2}}
\put(16,20){\oval(6,3)\makebox(0,0){$\scriptstyle\state{x}\mapsto\vecconjugate{\state{x}}$}}
\put(16,18.5){\vector(0,-1){3}}
\put(16,17.5){\line(-1,0){7}}

\put(16,14){\oval(4,3)\makebox(0,0){$L$}}
\put(20,14){\makebox(0,0){ $\scriptstyle L(\vecconjugate{\state{x}})$}}
\put(16,12.5){\vector(0,-1){3}}
\put(16,8.45){\makebox(0,0){$\vdots$}}
\put(16,6){\vector(0,-1){3}}
\put(16,1.5){\oval(4,3)\makebox(0,0){$L$}}
\put(21,1.5){\makebox(0,0){ $\scriptstyle L^{{\scriptscriptstyle3\len}}(\vecconjugate{\state{x}})$}}
\put(14,1.5){\vector(-1,0){3}}
\put(9,17.5){\vector(0,-1){14.25}}
\put(9,1.5){\oval(4,3.5)\makebox(0,0){$\IsEqual{?}$}}
\put(11,-1.5){\makebox(0,0){\small output}}
}

\newsavebox{\myCMPCL}
\savebox{\myCMPCL}(0,0){
\put(16,1.5){\oval(4,3)\makebox(0,0){$L$}}
\put(16,6){\vector(0,-1){3}}
\put(18,1.5){\line(1,0){6.5}}
\put(24.5,1.5){\line(0,-1){1.55}}
}

\newsavebox{\myCMPCLM} 
\savebox{\myCMPCLM}(0,0){
\put(16,1.5){\oval(4,3)\makebox(0,0){$L$}}
\put(16,6){\vector(0,-1){3}}
\put(18,1.5){\vector(1,0){3}}
\put(24.5,1.5){\oval(7,3)\makebox(0,0){$\min$}}
\put(24.5,6){\vector(0,-1){3}}
}

\newsavebox{\myCMPcircuit} 
\savebox{\myCMPcircuit}(0,0){ 
\put(16,39.5){\makebox(0,0){\small input $\state{x}$ }}
\put(16,38){\circle*{0.55}}
\put(16,38){\vector(0,-1){2}}
\put(16,34.5){\oval(6,3)\makebox(0,0){$\scriptstyle\state{x}\mapsto\vecconjugate{\state{x}}$}}
\put(0,30){\usebox{\myCMPCL}}
\put(16,26.9){\makebox(0,0)[tl]{ $\scriptstyle L\left(\vecconjugate{\state{x}}\right)$}}
\put(0,24){\usebox{\myCMPCLM}}
\put(16,20.9){\makebox(0,0)[tl]{
 $\scriptstyle L^{\scriptscriptstyle 2}\left(\vecconjugate{\state{x}}\right)$}}
\put(16,17.45){\makebox(0,0){$\vdots$}}
\put(16,21.1){\vector(0,-1){3}}
\put(24.5,17.45){\makebox(0,0){$\vdots$}}
\put(24.5,21.1){\vector(0,-1){3}}
\put(0,12){\usebox{\myCMPCLM}}
\put(16,8.9){\makebox(0,0)[tl]{ 
 $\scriptstyle L^{{\scriptscriptstyle3\len}}\left(\vecconjugate{\state{x}}\right)$}}
\put(16,5.45){\makebox(0,0){$\vdots$}}
\put(16,9.1){\vector(0,-1){3}}
\put(24.5,5.45){\makebox(0,0){$\vdots$}}
\put(24.5,9.1){\vector(0,-1){3}}
\put(0,0){\usebox{\myCMPCLM}}
\put(16,-3){\makebox(0,0)[tl]{ $\scriptstyle L^{\scriptscriptstyle 6\len}\left(\vecconjugate{\state{x}}\right)$}}
\put(3.5,8){\oval(4,3.5)\makebox(0,0){$\IsEqual{?}$}}
\put(16,8){\vector(-1,0){10.75}}
\put(16,32){\line(-1,0){12.5}}
\put(3.5,32){\vector(0,-1){22.25}}
\put(3.5,6.25){\vector(0,-1){12.75}}
\put(16,-3){\vector(0,-1){3.25}}
\put(16,-8){\oval(4,3.5)\makebox(0,0){$\IsEqual{?}$}}
\put(14,-8){\vector(-1,0){3}}
\put(9.5,-8){\circle{3}}
\put(9.5,-8){\makebox(0,0){$\myNOT$}}
\put(8,-8){\vector(-1,0){3}}
\put(3.5,-8){\circle{3}}
\put(3.5,-8){\makebox(0,0){$\myOR$}}
\put(24.5,-8){\vector(-1,0){6.5}}
\put(24.5,-3){\line(0,-1){5}}
\put(4.75,-10.5){\makebox(0,0){\small output}}
}

\newsavebox{\mySMPCL}
\savebox{\mySMPCL}(0,0){
\put(6,1.5){\line(-1,0){4.5}}
\put(1.5,1.5){\line(0,-1){1.55}}
\put(8.5,1.5){\oval(5,3.4)}
\put(14,1.5){\vector(-1,0){3}}
\put(16,1.5){\oval(4,3)\makebox(0,0){$L$}}
\put(16,6){\vector(0,-1){3}}
\put(18,1.5){\line(1,0){6.5}}
\put(24.5,1.5){\line(0,-1){1.55}}
}

\newsavebox{\mySMPCLM} 
\savebox{\mySMPCLM}(0,0){
\put(1.5,1.5){\circle{3}}
\put(1.5,1.5){\makebox(0,0){$\myOR$}}
\put(1.5,6){\vector(0,-1){3}}
\put(6,1.5){\vector(-1,0){3}}
\put(8.5,1.5){\oval(5,3.4)
}
\put(14,1.5){\vector(-1,0){3}}
\put(16,1.5){\oval(4,3)\makebox(0,0){$L$}}
\put(16,6){\vector(0,-1){3}}
\put(18,1.5){\vector(1,0){3}}
\put(24.5,1.5){\oval(7,3)\makebox(0,0){$\min$}}
\put(24.5,6){\vector(0,-1){3}}
}

\newsavebox{\myPSCL} 
\savebox{\myPSCL}(0,0){
\put(6,1.5){\line(-1,0){4.5}}
\put(1.5,1.5){\line(0,-1){1.55}}
\put(8.5,1.5){\oval(5,3.4)}
\put(14,1.5){\vector(-1,0){3}}
\put(16,1.5){\oval(4,3)\makebox(0,0){$L$}}
\put(16,6){\vector(0,-1){3}}
}

\newsavebox{\myPSCLM} 
\savebox{\myPSCLM}(0,0){
\put(1.5,1.5){\circle{3}}
\put(1.5,1.5){\makebox(0,0){$\myOR$}}
\put(1.5,6){\vector(0,-1){3}}
\put(6,1.5){\vector(-1,0){3}}
\put(8.5,1.5){\oval(5,3.4)
}
\put(14,1.5){\vector(-1,0){3}}
\put(16,1.5){\oval(4,3)\makebox(0,0){$L$}}
\put(16,6){\vector(0,-1){3}}
}

\newsavebox{\myPScircuit} 
\savebox{\myPScircuit}(0,0){
\put(16,34.5){\makebox(0,0){\small input $\state{x}$ }}
\put(16,33){\circle*{0.55}}
\put(0,30){\usebox{\myPSCL}}
\put(8.5,28.5){\makebox(0,0){$\crct{f_0}$}}
\put(16,26.9){\makebox(0,0)[tl]{ $\scriptstyle L(\state{x})$}}
\put(0,24){\usebox{\myPSCLM}}
\put(8.5,22.5){\makebox(0,0){$\crct{f_0}$}}
\put(16,20.9){\makebox(0,0)[tl]{
 $\scriptstyle L^{\scriptscriptstyle 2}(\state{x})$}}
\put(1.5,17.45){\makebox(0,0){$\vdots$}}
\put(1.5,21.1){\vector(0,-1){3}}
\put(8.5,17.45){\makebox(0,0){$\vdots$}}
\put(16,17.45){\makebox(0,0){$\vdots$}}
\put(16,21.1){\vector(0,-1){3}}
\put(0,12){\usebox{\myPSCLM}}
\put(8.5,10.5){\makebox(0,0){$\crct{f_0}$}}
\put(16,8.9){\makebox(0,0)[tl]{ 
 $\scriptstyle L^{{\scriptscriptstyle3\len}}(\state{x})$}}
\put(1.5,5.45){\makebox(0,0){$\vdots$}}
\put(1.5,9.1){\vector(0,-1){3}}
\put(8.5,5.45){\makebox(0,0){$\vdots$}}
\put(16,5.45){\makebox(0,0){$\vdots$}}
\put(16,9.1){\vector(0,-1){3}}
\put(0,0){\usebox{\myPSCLM}}
\put(8.5,-1.5){\makebox(0,0){$\crct{f_0}$}}
\put(16,-3){\makebox(0,0)[tl]{ $\scriptstyle L^{\scriptscriptstyle 6\len}(\state{x})$}}
\put(26,10.5){\oval(4,3.5)\makebox(0,0){$\IsEqual{?}$}}
\put(18,10.5){\vector(1,0){6}}
\put(16,33){\line(1,0){10}}
\put(26,33){\vector(0,-1){20.75}}
\put(26,8.75){\vector(0,-1){5.5}}
\put(26,2){\makebox(0,0){\small $p(\state{x})$}}
\put(1.5,-2.9){\vector(0,-1){2.55}}
\put(1.5,-7){\makebox(0,0){\small $s(\state{x})$}}
}
\newsavebox{\myMQCL}
\savebox{\myMQCL}(0,0){
\put(5,1.5){\line(-1,0){3.5}}
\put(1.5,1.5){\line(0,-1){1.55}}
\put(5,0){\framebox(6.5,3.0){CMP}}
\put(14,1.5){\vector(-1,0){2.5}}
\put(16,1.5){\oval(4,3)\makebox(0,0){$L$}}
\put(16,6){\vector(0,-1){3}}
\put(18,1.5){\line(1,0){6.5}}
\put(24.5,1.5){\line(0,-1){1.55}}
}

\newsavebox{\myMQCLM} 
\savebox{\myMQCLM}(0,0){
\put(1.5,1.5){\circle{3}}
\put(1.5,1.5){\makebox(0,0){$\myAND$}}
\put(1.5,6){\vector(0,-1){3}}
\put(5,1.5){\vector(-1,0){2}}
\put(5,0.0){\framebox(6.5,3.0){CMP}}
\put(14,1.5){\vector(-1,0){2.5}}
\put(16,1.5){\oval(4,3)\makebox(0,0){$L$}}
\put(16,6){\vector(0,-1){3}}
\put(18,1.5){\vector(1,0){3}}
\put(24.5,1.5){\oval(7,3)\makebox(0,0){$\min$}}
\put(24.5,6){\vector(0,-1){3}}
}

\newsavebox{\myMQcircuit} 
\savebox{\myMQcircuit}(0,0){
\put(16,34.5){\makebox(0,0){\small input $\state{x}$ }}
\put(16,33){\circle*{0.55}}
\put(0,30){\usebox{\myMQCL}}
\put(16,27.2){\makebox(0,0)[tr]{ $\scriptstyle L(\state{x})$}}
\put(0,24){\usebox{\myMQCLM}}
\put(16,20.9){\makebox(0,0)[tr]{
 $\scriptstyle L^{\scriptscriptstyle 2}(\state{x})$}}
\put(1.5,17.45){\makebox(0,0){$\vdots$}}
\put(1.5,21.1){\vector(0,-1){3}}
\put(8.5,17.45){\makebox(0,0){$\vdots$}}
\put(16,17.45){\makebox(0,0){$\vdots$}}
\put(16,21.1){\vector(0,-1){3}}
\put(24.5,17.45){\makebox(0,0){$\vdots$}}
\put(24.5,21.1){\vector(0,-1){3}}
\put(0,12){\usebox{\myMQCLM}}
\put(16,8.6){\makebox(0,0)[tr]{ 
 $\scriptstyle L^{{\scriptscriptstyle\len}}(\state{x})$}}
\put(1.5,5.45){\makebox(0,0){$\vdots$}}
\put(1.5,9.1){\vector(0,-1){3}}
\put(8.5,5.45){\makebox(0,0){$\vdots$}}
\put(16,5.45){\makebox(0,0){$\vdots$}}
\put(16,9.1){\vector(0,-1){3}}
\put(24.5,5.45){\makebox(0,0){$\vdots$}}
\put(24.5,9.1){\vector(0,-1){3}}
\put(0,0){\usebox{\myMQCLM}}
\put(16,-3.3){\makebox(0,0)[tr]{ $\scriptstyle L^{\scriptscriptstyle 6\len}(\state{x})$}}
\put(30,-8){\oval(4,3.5)\makebox(0,0){$\IsEqual{?}$}}
\put(16,8){\line(1,0){8.3}}
\put(24.5,8){\oval(0.05,0.05)[t]}
\put(24.7,8){\line(1,0){5.3}}
\put(30,8){\vector(0,-1){14.25}}
\put(16,-3){\vector(0,-1){3.25}}
\put(16,-8){\oval(4,3.5)\makebox(0,0){$\IsEqual{?}$}}
\put(24.5,-8){\vector(-1,0){6.5}}
\put(24.5,-8){\vector(1,0){3.5}}
\put(24.5,-3){\vector(0,-1){7}}
\put(22,-13){\framebox(5,3){CP}}
\put(24.5,-13){\vector(0,-1){2}}
\put(24.5,-16.5){\circle{3}}
\put(24.5,-16.5){\makebox(0,0){$\myAND$}}
\put(30,-9.75){\line(0,-1){6.75}}
\put(30,-16.5){\vector(-1,0){4}}
\put(8,-16.5){\circle{3}}
\put(8,-16.5){\makebox(0,0){$\myAND$}}
\put(6.5,-16.5){\vector(-1,0){2}}
\put(23,-16.5){\vector(-1,0){13.5}}
\put(16,-9.75){\vector(0,-1){2.5}}
\put(16,-13.5){\makebox(0,0){\small $m(\state{x})$}}
\put(1.5,-2.9){\line(0,-1){5.1}}
\put(1.5,-8){\line(1,0){6.5}}
\put(8,-8){\vector(0,-1){7}}
\put(2.5,-15.5){\makebox(0,0){\small $q(\state{x})$}}
}

\newsavebox{\myREDcircuit}
\savebox{\myREDcircuit}(0,0){ 
\put(0,25){\makebox(0,0)[b]{\scriptsize input $\state{x}$}}
\put(0,24){\line(0,-1){1.5}}
\put(0,24){\circle*{0.55}}
\put(0,22.5){\line(1,0){22}}
\put(0,22.5){\line(-1,0){22}}
\put(22,22.5){\line(0,-1){19}}
\put(22,19.5){\vector(-1,0){4}}
\put(22,3.5){\vector(-1,0){4}}
\put(11.5,22.5){\makebox(0,0)[bl]{$\scriptstyle\state{z}=1\parallel\state{x}$}}
\put(-10.5,22.5){\makebox(0,0)[br]{$\scriptstyle\state{y}=0\parallel\state{x}$}}
\put(-22,22.5){\line(0,-1){19}}
\put(-22,19.5){\vector(1,0){4}}
\put(-22,3.5){\vector(1,0){4}}
\put(-12,3.5){\vector(0,1){4.5}}
\put(-12,9.5){\circle{3}}
\put(-12,9.5){\makebox(0,0){$\myAND$}}
\put(-12,11){\vector(0,1){2}}
\put(-12,14.5){\circle{3}}
\put(-12,14.5){\makebox(0,0){$\myAND$}}
\put(-12,19.5){\vector(0,-1){3.5}}
\put(-18,2){\framebox(4,3){\scriptsize MQ}}
\put(-14,19.5){\vector(1,0){6.5}}
\put(-14,3.5){\vector(1,0){6.5}}
\put(-18,18){\framebox(4,3){\scriptsize PS}}
\put(-16,18){\line(0,-1){8.5}}
\put(-16,9.5){\vector(1,0){2.5}}
\put(-16,0){\line(1,0){10}}
\put(-16,2){\line(0,-1){2}}
\put(-4.5,19.5){\vector(1,0){3}}
\put(-6,19.5){\circle{3}}
\put(-6,19.5){\makebox(0,0){$\myNOT$}}
\put(-6,3.5){\circle{3}}
\put(-6,3.5){\makebox(0,0){$\myOR$}}
\put(-6,0){\vector(0,1){2}}
\put(-4.5,3.5){\vector(1,0){3}}
\put(0,3.5){\circle{3}}
\put(0,3.5){\makebox(0,0){$\myOR$}}
\put(0,14.5){\circle{3}}
\put(0,14.5){\makebox(0,0){$\myAND$}}
\put(0,5){\vector(0,1){8}}
\put(0,18){\vector(0,-1){2}}
\put(0,19.5){\circle{3}}
\put(0,19.5){\makebox(0,0){$\myAND$}}
\put(4.5,3.5){\vector(-1,0){3}}
\put(6,0){\vector(0,1){2}}
\put(6,3.5){\circle{3}}
\put(6,3.5){\makebox(0,0){$\myOR$}}
\put(4.5,19.5){\vector(-1,0){3}}
\put(6,19.5){\circle{3}}
\put(6,19.5){\makebox(0,0){$\myNOT$}}
\put(12,3.5){\vector(0,1){4.5}}
\put(12,9.5){\circle{3}}
\put(12,9.5){\makebox(0,0){$\myAND$}}
\put(12,11){\vector(0,1){2}}
\put(12,14.5){\circle{3}}
\put(12,14.5){\makebox(0,0){$\myAND$}}
\put(12,19.5){\vector(0,-1){3.5}}
\put(14,2){\framebox(4,3){\scriptsize MQ}}
\put(14,19.5){\vector(-1,0){6.5}}
\put(14,3.5){\vector(-1,0){6.5}}
\put(14,18){\framebox(4,3){\scriptsize PS}}
\put(16,18){\line(0,-1){8.5}}
\put(16,9.5){\vector(-1,0){2.5}}
\put(16,0){\line(-1,0){10}}
\put(16,2){\line(0,-1){2}}
\put(-1.5,14.5){\vector(-1,0){3}}
\put(-6,14.5){\circle{3}}
\put(-6,14.5){\makebox(0,0){$\myOR$}}
\put(-10.5,14.5){\vector(1,0){3}}
\put(-6,13){\line(0,-1){2}}
\put(10.5,14.5){\line(-1,0){4.5}}
\put(6,14.5){\vector(0,-1){2}}
\put(6,11){\circle{3}}
\put(6,11){\makebox(0,0){$\myOR$}}
\put(6,8){\makebox(0,0){\scriptsize output}}
\put(-6,11){\line(1,0){5.85}}
\put(0,11){\oval(0.05,0.05)[t]}
\put(0.1,11){\vector(1,0){4.4}}
\put(13.5,19.5){\makebox(0,0)[br]{$\scriptstyle p(\state{z})$}}
\put(13.5,3.5){\makebox(0,0)[tr]{$\scriptstyle m(\state{z})$}}
\put(16.3,0){\makebox(0,0)[bl]{$\scriptstyle q(\state{z})$}}
\put(16.3,16){\makebox(0,0)[tl]{$\scriptstyle s(\state{z})$}}
\put(-13.5,19.5){\makebox(0,0)[bl]{$\scriptstyle p(\state{y})$}}
\put(-13.5,3.5){\makebox(0,0)[tl]{$\scriptstyle m(\state{y})$}}
\put(-16,0){\makebox(0,0)[br]{$\scriptstyle q(\state{y})$}}
\put(-15.8,16){\makebox(0,0)[tr]{$\scriptstyle s(\state{y})$}}
}

\newsavebox{\myDECcircuit} 
\savebox{\myDECcircuit}(0,0){ 
\put(16,28){\makebox(0,0){\small input $\state{x}$ }}
\put(16,26.75){\circle*{0.55}}
\put(16,26.75){\vector(0,-1){2.5}}
\put(16,22.75){\oval(4,3)\makebox(0,0){$M$}}
\put(16,21){\makebox(0,0)[tr]{ $\scriptstyle \state{u}_0$}}
\put(0,18){\usebox{\mySMPCL}}
\put(8.5,16.5){\makebox(0,0){$\crct{f_2}$}}
\put(16,14.9){\makebox(0,0)[tr]{ $\scriptstyle L(\state{u}_0)$}}
\put(0,12){\usebox{\mySMPCLM}}
\put(8.5,10.5){\makebox(0,0){$\crct{f_2}$}}
\put(16,8.9){\makebox(0,0)[tr]{
 $\scriptstyle L^{\scriptscriptstyle 2}(\state{u}_0)$}}
 \put(1.5,5.45){\makebox(0,0){$\vdots$}}
\put(1.5,9.1){\vector(0,-1){3}}
\put(8.5,5.45){\makebox(0,0){$\vdots$}}
\put(16,5.45){\makebox(0,0){$\vdots$}}
\put(16,9.1){\vector(0,-1){3}}
\put(24.5,5.45){\makebox(0,0){$\vdots$}}
\put(24.5,9.1){\vector(0,-1){3}}
\put(0,0){\usebox{\mySMPCLM}}
\put(8.5,-1.5){\makebox(0,0){$\crct{f_2}$}}
\put(16,-3.25){\makebox(0,0)[tr]{ $\scriptstyle L^{\scriptscriptstyle 3\len}(\state{u}_0)$}}
\put(16,-3){\vector(0,-1){3.25}}
\put(16,-8){\oval(4,3.5)\makebox(0,0){$\IsEqual{?}$}}
\put(24.5,-8){\vector(-1,0){6.5}}
\put(24.5,-3){\line(0,-1){5}}
\put(14,-8){\vector(-1,0){4}}
\put(1.5,-2.9){\line(0,-1){5.1}}
\put(1.5,-8){\vector(1,0){5.5}}
\put(8.5,-8){\circle{3}}
\put(8.5,-8){\makebox(0,0){$\myAND$}}
\put(8.5,-9.5){\makebox(0,0)[t]{\small output}}
}

\newsavebox{\myCnull}
\savebox{\myCnull}(0,0){ 
\put(0.5,33.25){\vector(0,-1){14}}
\put(0.5,17.5){\oval(6,3.5)\makebox(0,0){$\myAND_r$}}
\put(0.5,15.75){\line(0,-1){14.25}}
\put(0.5,1.5){\vector(1,0){5.5}}
\put(0.5,7){\vector(1,0){2.5}}
\put(0.5,33.25){\line(1,0){21.5}}
\put(4.5,7){\circle{3}}
\put(4.5,7){\makebox(0,0){$\myNOT$}}
\put(6,7){\line(1,0){1.25}}
\put(7.5,7){\oval(0.05,0.05)[t]}
\put(7.75,7){\vector(1,0){7.75}}
\put(7.5,33.25){\vector(0,-1){2.5}}
\put(7.5,29.25){\oval(4,3)}
\put(7.5,29.25){\makebox(0,0){$\myNOT_r$}}
\put(7.5,27.75){\vector(0,-1){2.5}}
\put(7.5,23.5){\oval(5,3.5)}
\put(7.5,23.5){\makebox(0,0){$\crct{f_0}$}}
\put(7.5,21.75){\vector(0,-1){2.25}}
\put(7.5,18){\circle{3}}
\put(7.5,18){\makebox(0,0){$\myAND$}}
\put(7.5,16.5){\vector(0,-1){2.5}}
\put(7.5,12.5){\circle{3}}
\put(7.5,12.5){\makebox(0,0){$\myXOR$}}
\put(7.5,11){\vector(0,-1){8}}
\put(7.5,1.5){\circle{3}}
\put(7.5,1.5){\makebox(0,0){$\myAND$}}
\put(9.0,1.5){\vector(1,0){6.5}}
\put(13,18){\circle{3}}
\put(13,18){\makebox(0,0){$\myNOT$}}
\put(11.5,18){\vector(-1,0){2.5}}
\put(17,18){\vector(-1,0){2.5}}
\put(10.5,38){\makebox(0,0){\scriptsize input}}
\put(10.5,36.25){\line(0,-1){3}}
\put(10.5,36.25){\circle*{1}}
\put(17,33.25){\vector(0,-1){8}}
\put(17,23.5){\oval(5,3.5)}
\put(17,23.5){\makebox(0,0){$\crct{f_0}$}}
\put(17,21.75){\vector(0,-1){13.25}}
\put(17,12.5){\vector(-1,0){8}}
\put(17,7){\circle{3}}
\put(17,7){\makebox(0,0){$\myAND$}}
\put(17,5.5){\vector(0,-1){2.5}}
\put(17,1.5){\circle{3}}
\put(17,1.5){\makebox(0,0){$\myXOR$}}
\put(18.5,1.5){\vector(1,0){2}}
\put(22,33.25){\vector(0,-1){2.5}}
\put(22,29.25){\oval(4,3)}
\put(22,29.25){\makebox(0,0){$\myNOT_r$}}
\put(22,27.75){\vector(0,-1){8.5}}
\put(22,17.5){\oval(6,3.5)\makebox(0,0){$\myAND_r$}}
\put(22,15.75){\vector(0,-1){7.25}}
\put(22,7){\circle{3}}
\put(22,7){\makebox(0,0){$\myNOT$}}
\put(22,5.5){\vector(0,-1){2.5}}
\put(22,1.5){\circle{3}}
\put(22,1.5){\makebox(0,0){$\myAND$}}
\put(22,-0.5){\makebox(0,0)[t]{\scriptsize output}}
}

\newsavebox{\myLFSRST}
\savebox{\myLFSRST}(0,0){ 
\put(0,-3){\makebox(0,0)[t]{{$\scriptstyle y_0$}}}
\put(4.5,-3){\makebox(0,0)[t]{$\scriptstyle y_1$}}
\put(13.5,-3){\makebox(0,0)[t]{$\scriptstyle y_{m-1}$}}
\put(18,-3){\makebox(0,0)[t]{$\scriptstyle y_{m}$}}
\put(22.5,-3){\makebox(0,0)[t]{$\scriptstyle y_{m+1}$}}
\put(31.5,-3){\makebox(0,0)[t]{$\scriptstyle y_{2m-2}$}}
\put(36,-1.5){\makebox(0,0)[t]{$\scriptstyle y_{2m-1}$}}
\multiput(0,0)(4.5,0){2}{\vector(0,-1){3}}
\put(9,-1.5){\makebox(0,0){$\dots$}}
\multiput(13.5,0)(4.5,0){3}{\vector(0,-1){3}}
\put(27,-1.5){\makebox(0,0){$\dots$}}
\put(31.5,0){\vector(0,-1){3}}
\put(36,0){\vector(0,-1){1.5}}
\put(18,9){\circle*{0.45}}
\put(18,9){\vector(0,-1){3.75}}
\put(18,4){\circle{2.5}}
\put(18,4){\makebox(0,0){$\myXOR$}}
\put(19.25,4){\line(1,0){16.75}}
\put(36,4){\line(0,-1){4}}
\multiput(4.5,9)(4.5,0){2}{\line(-1,-2){4.5}}
\multiput(18,9)(4.5,0){3}{\line(-1,-2){4.5}}
\put(36,9){\line(-1,-2){4.5}}
\put(12,6.5){\makebox(0,0){$\dots$}}
\put(30,6.5){\makebox(0,0){$\dots$}}
\put(0,9){\line(0,-1){5}}
\put(0,4){\vector(1,0){16.75}}
\multiput(0,12)(4.5,0){3}{\line(0,-1){3}}
\put(12,10.5){\makebox(0,0){$\dots$}}
\multiput(18,12)(4.5,0){3}{\line(0,-1){3}}
\put(31.5,10.5){\makebox(0,0){$\dots$}}
\put(36,12){\line(0,-1){3}}
\put(0,12){\makebox(0,0)[b]{{$\scriptstyle x_0$}}}
\put(4.5,12){\makebox(0,0)[b]{$\scriptstyle x_1$}}
\put(9,12){\makebox(0,0)[b]{$\scriptstyle x_{2}$}}
\put(18,12){\makebox(0,0)[b]{$\scriptstyle x_{m}$}}
\put(22.5,12){\makebox(0,0)[b]{$\scriptstyle x_{m+1}$}}
\put(27,12){\makebox(0,0)[b]{$\scriptstyle x_{m+2}$}}
\put(36,12){\makebox(0,0)[b]{$\scriptstyle x_{2m-1}$}}
}

\newsavebox{\myMIN}
\savebox{\myMIN}(0,0){  
\put(0,9.5){\makebox(0,0)[b]{\scriptsize input $\state{x}$}}
\put(-10,8.5){\makebox(0,0)[b]{\scriptsize $\hbit{m-1}{\state{x}}$}}
\put(7,8.5){\makebox(0,0)[b]{\scriptsize ${\state{x}}$}}
\put(0,6.5){\makebox(0,0)[r]{\scriptsize $x_{m-1}$}}
\put(0,8.5){\circle*{1}}
\put(0,8.5){\line(1,0){16}}
\put(0,8.5){\line(-1,0){23}}
\put(0,-8.5){\circle*{1}}
\put(0,-8.5){\line(1,0){16}}
\put(0,-8.5){\line(-1,0){9.9}}
\put(-10,-8.5){\oval(0.05,0.05)[t]}
\put(-23,-8.5){\line(1,0){12.9}}
\put(0,-9.5){\makebox(0,0)[t]{\scriptsize  input $\state{y}$}}
\put(0,-6.5){\makebox(0,0)[r]{\scriptsize $y_{m-1}$}}
\put(-17,-8){\makebox(0,0)[b]{\scriptsize $\hbit{m-1}{\state{y}}$}}
\put(7,-8.7){\makebox(0,0)[t]{\scriptsize ${\state{y}}$}}
\put(-12,-12.5){\makebox(0,0)[r]{\scriptsize output $\min_m(\state{x},\state{y})$}}
\put(0,0){\circle{3}}
\put(0,0){\makebox(0,0){$\myXOR$}}
\put(0,8){\vector(0,-1){6.5}}
\put(0,-8){\vector(0,1){6.5}}
\put(4,5){\circle{3}}
\put(0,5){\vector(1,0){2.5}}
\put(4,5){\makebox(0,0){$\myNOT$}}
\put(4,-5){\circle{3}}
\put(4,-5){\makebox(0,0){$\myNOT$}}
\put(0,-5){\vector(1,0){2.5}}
\put(10,5){\circle{3}}
\put(5.5,5){\vector(1,0){3}}
\put(10,5){\makebox(0,0){$\myAND$}}
\put(10,-5){\circle{3}}
\put(10,-5){\makebox(0,0){$\myAND$}}
\put(5.5,-5){\vector(1,0){3}}
\put(10,0){\vector(0,-1){3.5}}
\put(10,0){\vector(0,1){3.5}}
\put(0,5){\circle*{0.55}}
\put(0,-5){\circle*{0.55}}
\put(1.5,0){\line(1,0){8.5}}
\put(16,5){\circle{3}}
\put(16,5){\makebox(0,0){$\times$}}
\put(16,8.5){\vector(0,-1){2}}
\put(16,-5){\circle{3}}
\put(16,-5){\makebox(0,0){$\times$}}
\put(16,-8.5){\vector(0,1){2}}
\put(11.5,5){\vector(1,0){3}}
\put(11.5,-5){\vector(1,0){3}}
\put(16,0){\circle{3}}
\put(16,0){\makebox(0,0){$\myXOR$}}
\put(16,3.5){\vector(0,-1){2}}
\put(16,-3.5){\vector(0,1){2}}
\put(17.5,0){\line(1,0){1.5}}
\put(19,0){\line(0,-1){12}}
\put(-1.5,0){\vector(-1,0){2}}
\put(-5,0){\circle{3}}
\put(-5,0){\makebox(0,0){$\myNOT$}}
\put(-6.5,0){\vector(-1,0){2}}
\put(-10,0){\circle{3}}
\put(-10,0){\makebox(0,0){$\times$}}
\put(-10,-1.5){\vector(0,-1){9}}
\put(-10,-12){\circle{3}}
\put(-10,-12){\makebox(0,0){$\myXOR$}}
\put(19,-12){\vector(-1,0){27.5}}
\put(-13.5,0){\vector(1,0){2}}
\put(-13,0.5){\makebox(0,0){\scriptsize $z$}}
\put(-15,5){\line(1,0){15}}
\put(-15,5){\vector(0,-1){3.5}}
\put(-15,0){\circle{3}}
\put(-15,0){\makebox(0,0){$\parallel$}}
\put(-19,0){\vector(1,0){2.5}}
\put(-23,0){\oval(8,3)}
\put(-23,0){\makebox(0,0){\scriptsize $\min_{m-1}$}}
\put(-23,8.5){\vector(0,-1){7}}
\put(-23,-8.5){\vector(0,1){7}}
}

\begin{document}
\title{Deciding Irreducibility/Indecomposability
of
Feedback Shift Registers
is \NP-hard}
\author{
{Lin W{\small ANG}
}\\
{\it\small Science and Technology on Communication Security Laboratory}\\
{\it\small Chengdu 610041, P. R. China}\\
{\it\small Email: linwang@math.pku.edu.cn}}
\date{}%
\maketitle
\begin{abstract}
Feedback shift registers(FSRs) are 
a fundamental component 
in electronics and secure communication.
An FSR $f$ is said to be reducible if all the output sequences of
another FSR $g$ 
can also be generated by $f$
and the FSR $g$ has less memory than $f$.
An FSR is said to be decomposable if it has the same set of output sequences
as a cascade connection of two FSRs.
It is proved that deciding whether FSRs are irreducible/indecomposable
is \NP-hard.

\NP-hard,
Boolean circuit,
cycle structure

\end{abstract}

\section{Introduction}\label{sect:intro}
%
Feedback shift registers are broadly used in
spread spectrum radio, control engineering 
and confidential digital communication.
Consequently, this subject has attracted
substantial research over half a century.
Particularly, feedback shift registers
play a significant role in the stream cipher finalists
of the eSTREAM project\cite{RB08}.

\begin{figure}[htb]
\setlength{\unitlength}{1mm}
\begin{center}
\begin{picture}(80,10.5)
\put(0,0){\usebox{\myFSR}}
\put(9,-3){\makebox(0,0){$\seq{x_{n-1}}$}}
\put(22,-3){\makebox(0,0){$\seq{x_{n-2}}$}}
\put(48,-3){\makebox(0,0){$\seq{x_{1}}$}}
\put(61,-3){\makebox(0,0){$\seq{x_{0}}$}}
\put(35,5.5){\makebox(0,0){$\crct{f_1}(\seq{x_0},\seq{x_1},\dots,\seq{x_{n-1}})$}}
\put(70,-3){\makebox(0,0)[l]{output}}
\end{picture}
\end{center}
\caption{A feedback shift register with feedback logic $\crct{f_1}$}\label{fig:FSR}
\end{figure}
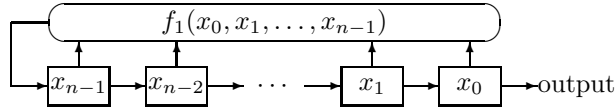
As shown in Figure \ref{fig:FSR},
an $n$-stage \emph{feedback shift register(FSR)} consists of
$n$ bit registers $\seq{x_0},\seq{x_1},\dots,\seq{x_{n-1}}$
and an $n$-input feedback logic $f_1$.
The vector $\left(\seq{x_0}(t),\seq{x_1}(t),\dots,\seq{x_{n-1}}(t)\right)$
is called a \emph{state} of this FSR,
where $\seq{x_i}(t)$ is the value of $\seq{x_i}$ at clock cycle $t$,
$0\leq i<n$.
The state at clock cycle $0$ is called the \emph{initial state}.
Along with clock impulses
the values stored in bit registers
update themselves as 
\begin{equation}\label{eqn:def-fsr-statetrans}
\left(\seq{x_0}(t+1),\seq{x_1}(t+1),\dots,\seq{x_{n-1}}(t+1)\right)=
\left(\seq{x_1}(t),\dots,\seq{x_{n-1}}(t),f_1(\seq{x_0}(t),\seq{x_1}(t),\dots,\seq{x_{n-1}}(t))\right),
\end{equation}
and the map defined by Eq.(\ref{eqn:def-fsr-statetrans})
is called the \emph{state transformation} of this FSR.

The $(n+1)$-input Boolean function
$f(x_0,x_1,\dots,x_{n}) = x_{n}\oplus f_1(x_0,x_1,\dots,x_{n-1})$,
where $\oplus$ denotes exclusive-or,
is called the characteristic function of the FSR in Figure \ref{fig:FSR},
and without ambiguity we also denote this FSR by $f$.
Let $\SeqSet{f}$ denote the set of sequences generated by 
$f$,
i.e.,
\begin{equation*}
\SeqSet{f}=\set{\seq{s}\in\BnrSet^*:\forall t, f(\seq{s}(t),\seq{s}(t+1),\dots,\seq{s}(t+n))=0},
\end{equation*}
where $\BnrSet^*$ is the set of binary sequences.
If $f(x_0,x_1,\dots,x_\len)= x_{\len}\oplus c_{\len-1}x_{\len-1}\oplus
\cdots \oplus c_1x_{1}\oplus c_0x_{0}$, where
$c_0,c_1,\dots,c_{\len-1}\in\BnrSet$, then 
$f$ is called a \emph{linear feedback shift register(LFSR)}, and
$p(x)=x^{\len}\oplus c_{\len-1}x^{\len-1}\oplus \cdots \oplus c_1x \oplus
c_0$ is called its \emph{characteristic polynomial}.
Without ambiguity we also denote this LFSR by $p(x)$.
An FSR which is not an LFSR is called a \emph{nonlinear feedback shift register(NFSR)}.

If there exists an $m$-stage FSR $g$
such that $m<n$ and $\SeqSet{g}\subset\SeqSet{f}$,
then $g$ is called a \emph{subFSR} of $f$ and
$f$ is said to be \emph{reducible}.
Otherwise, $f$ is said to be \emph{irreducible}.

\begin{figure}[htb]
\begin{center}
\setlength{\unitlength}{1mm}
\begin{picture}(150,10.5)
\put(0,0){\usebox{\myFSR}}
\put(9,-3){\makebox(0,0){$x_{n-1}$}}
\put(22,-3){\makebox(0,0){$x_{n-2}$}}
\put(48,-3){\makebox(0,0){$x_{1}$}}
\put(61,-3){\makebox(0,0){$x_{0}$}}
\put(35,5.5){\makebox(0,0){$\crct{f_1}(x_0,x_1,\dots,x_{n-1})$}}
\put(70,-3.05){\line(1,0){3}}
\put(71.5,-3.05){\circle{3}}
\put(71.45,-4.5){\line(0,1){3}}
\put(71.45,3){\vector(0,-1){4.9}}
\put(71.5,0){\usebox{\myFSR}}
\put(80.5,-3){\makebox(0,0){$y_{m-1}$}}
\put(93.5,-3){\makebox(0,0){$y_{m-2}$}}
\put(119.5,-3){\makebox(0,0){$y_{1}$}}
\put(132.5,-3){\makebox(0,0){$y_{0}$}}
\put(106.5,5.5){\makebox(0,0){$\crct{g_1}(y_0,y_1,\dots,y_{m-1})$}}
\put(141.5,-3){\makebox(0,0)[l]{output}}
\end{picture}
\end{center}
\caption{The cascade connection of $f$ in $g$}\label{fig:cascade}
\end{figure}
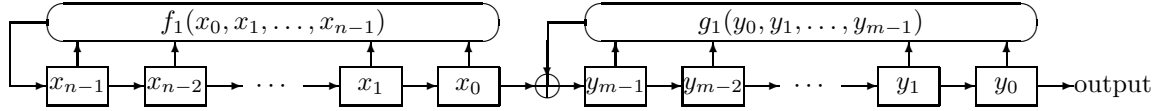

Let $f(x_0,x_1,\dots,x_{n})= x_n \oplus \crct{f_1}(x_0,x_1,\dots,x_{n-1})$
and $g(y_0,y_1,\dots,y_{m})= y_m \oplus \crct{g_1}(y_0,y_1,\dots,y_{m-1})$
be two FSRs.
The finite state machine in Figure \ref{fig:cascade}
is called the \emph{cascade connection} of $f$ into $g$.
The Grain family ciphers use the cascade connection of
an LFSR into an NFSR\cite{HJMM08}.
Green and Dimond\cite{GD70}
defined the \emph{product FSR}\footnote{
The product FSR of $f$ and $g$ is denoted
by $f.g$ in \cite{GD70}, while by $f\cascade g$ in \cite{MST79}.
We follow the latter in order to avoid ambiguity with periods or conventional multiplication.}
of $f$ and $g$ to be
\begin{equation*}
(\crct{f}\cascade\crct{g})(x_0,x_1,\dots,x_{n+m})=
\crct{f}(\crct{g}(x_0,x_1,\dots,x_{m}),
\crct{g}(x_1,x_2,\dots,x_{m+1}),\dots,\crct{g}(x_{n},x_{n+1},\dots,x_{n+m})),
\end{equation*}
and showed $\SeqSet{f;g}=\SeqSet{f\cascade g}$,
where $\SeqSet{f;g}$ is the set of output sequences of 
the cascade connection of $f$ into $g$.
Given an FSR $h$,
if there exist two FSRs $f$ and $g$
satisfying  $h=f\cascade g$,
then $h$ is said to be \emph{decomposable}.
Otherwise, $h$ is said to be \emph{indecomposable}.

It is appealing to decide whether an FSR is (ir)reducible/(in)decomposable
for the reasons below.
First, it offers a new perspective on analysis of stream ciphers.
Notice that all sequences generated by $g$ is also generated by
$f\cascade g$ if $f$ can output the 0-sequence.
A reducible/decomposable FSR in unaware use may undermine
the claimed security of stream ciphers, e.g.,
causing inadequate period of the output sequences.
Particularly, if $g$ is an LFSR and
$f$ can output the 0-sequence, then $f\cascade g$ can
generate a family of linear recurring sequences,
vulnerable to the Berlekamp-Massey algorithm.
Second, it potentially improves implementation of FSRs.
On one hand, it costs less memory to
replace an FSR with its large-stage subFSR, if there is one,
while generating a great part of its output sequences.
On the other hand, similar to the idea of Dubrova\cite{dbrv},
substituting a decomposable FSR by its equivalent cascade connection as in Figure \ref{fig:cascade}
possibly reduces the circuit depth of the feedback logics,
in favor of less propagation time and larger throughput.
Third, an algorithm testing (ir)reducibility/(in)decomposability
 helps to design useful FSRs.
Because Tian and Qi\cite{TQ13} proved that on average
at least one among three randomly chosen NFSRs is irreducible,
a great number of irreducible NFSRs can be found
if deciding irreducibility of FSRs is feasible.
Besides, FSRs generating maximal-length sequences were
constructed based on inherent structure of decomposable FSRs\cite{MST79}.

Two algorithms were proposed by \cite{TQ14}
to find affine subFSRs of NFSRs.
By \cite{JL16},
if an NFSR $h$ is decomposed as the
cascade connection of an LFSR $f$ into an NFSR $g$
and $f$ is primitive with stage no less than that of $g$,
then all affine subFSRs of $h$
are actually those 
of $g$.
(In)decomposability of LFSRs
is completely determined by their characteristic polynomials.
By \cite{GD70,LN83,TQ14s},
an LFSR $h$, with its characteristic polynomial $p(x)$,
is decomposed as $h=f\cascade g$ if and only if
$f$ and $g$ are LFSRs and $p(x)=l_1(x)\cdot l_2(x)$,
where $l_1(x)$ and $l_2(x)$  are
characteristic polynomials of $f$ and $g$, respectively.
In contrast, decomposing NFSRs seems much more challenging,
though some progress has been made recently.
Using the language of algebraic normal forms of Boolean functions,
Ma \emph{et al}\cite{MQT13} gave an algorithm to
 decompose NFSRs into the cascade connection
of an NFSR into an LFSR,
and Tian and Qi\cite{TQ14s} gave a series of algorithms to
decompose NFSRs into the cascade connection
of two NFSRs.
Noteworthily, Zhang \emph{et al}\cite{ZQTW15} gained
an algorithm decomposing an NFSR $f$ into the cascade connection
of an NFSR into an LFSR, and the complexity of their algorithm is
polynomial in the size of the algebraic normal form of $f$
and the size of the binary decision diagram of $f$ if
converting the algebraic normal form of $f$ to
the binary decision diagram of $f$ is polynomial-time computable.

\emph{Our contribution.} This correspondence studies irreducibility and indecomposability
from the perspective of computational complexity.
\NP\, is the class of all problems computed by polynomial-time nondeterministic Turing machines.
A problem is \NP-\emph{hard} if it is at least as hard as all \NP\, problems.
This correspondence proves that
deciding whether an FSR is irreducible(indecomposable)
is \NP-hard.

The rest of this paper is organized as follows: In Section
\ref{sect:preliminaries}
we prepare some notations, basic facts on Boolean circuits
and some lemmas on the cycle structure of FSRs.
\NP-hardness of FSR irreducibility and
FSR indecomposability is shown in Sections
\ref{sect:FSR-irreducibility} and \ref{sect:FSR-indecomposability}, respectively.
The last section includes a summary and a comment on future work.

\section{Preliminaries}\label{sect:preliminaries}

\subsection{Notations}

Throughout this paper,
$\Int$ denotes the set of integers,
\textquotedblleft$+$\textquotedblright\, addition of integers,
and \textquotedblleft$\oplus$\textquotedblright\, the exclusive-or(XOR) operation.

Denote
$\vecone{m}=(1,1,\dots,1)\in\BnrSet^{m}$,
$\vecnull{m}=(0,0,\dots,0)\in\BnrSet^{m}$
and $\veconenull{m}=(1,0,\dots,0)\in\BnrSet^{m}$.
For $\state{u}\in\BnrSet^m$,
denote $\veccomplement{\state{u}}=\state{u}\oplus\vecone{m}$
and $\vecconjugate{\state{u}}=\state{u}\oplus\veconenull{m}$.

For $\state{u}=(a_1,a_2,\dots,a_m)\in\BnrSet^m$
and $1\leq k<m$,
let
\begin{align*}
\hbit{k}{\state{u}}=&(a_1,a_2,\dots,a_k)\in\BnrSet^k;\\
\lbit{k}{\state{u}}=&(a_{m-k+1},a_{m-k+2},\dots,a_{m-1},a_m)\in\BnrSet^k.
\end{align*}

For $\state{u}=(a_1,\dots,a_k)\in\BnrSet^k$
and $\state{v}=(b_1,\dots,b_m)\in\BnrSet^m$,
denote $\state{u}\parallel\state{v}=(a_1,\dots,a_k,b_1,\dots,b_m)\in\BnrSet^{k+m}$.

Without ambiguity
a vector $(a_0,a_1,\dots,a_{m-1})\in\BnrSet^{m}$
is uniquely taken as 
the nonnegative integer $\sum_{j=0}^{m-1}2^{j}a_j$.
Thereby, the natural order relation on $\BnrSet^{m}$ is imposed, i.e.,
$(a_0,a_1,\dots,a_{m-1})<(b_0,b_1,\dots,b_{m-1})$
  if and only if
 $\sum_{j=0}^{m-1}2^{j}a_j<\sum_{j=0}^{m-1}2^{j}b_j$.

\subsection{Boolean circuits}

An $m$-input \emph{Boolean circuit} $f$ is a
directed acyclic graph with $m$ sources
and one sink
\cite{AB12}.
The value(s) of source(s) is(are) input(s) of the Boolean circuit;
Any nonsource vertex, called a \emph{gate}, is one of
the logical operations  OR($\myNOT$), AND($\myAND$) and  NOT($\myNOT$),
where the fan-in\footnote{The fan-in of a gate is the number of 
bits fed into it.} of OR and AND is $2$ and that of NOT is $1$;
The value outputted from a gate is obtained by applying its logical operation on the value(s) inputted into it;
The value outputted from the sink is the output of the Boolean circuit $f$.
The size of the circuit $f$,
denoted by $\sizeof{f}$, is  the number of vertices in it.
An $m$-input Boolean circuit $f$ is \emph{satisfiable} if
there exists $\state{v}\in\BnrSet^m$ such that
$f(\state{v})=1$.

\begin{center}
\begin{minipage}[h]{0.7\textwidth}
\textbf{PROBLEM}: CIRCUIT SATISFIABILITY

{INSTANCE}: A Boolean circuit $f$ with its size $\sizeof{f}$.

{QUESTION}: Is $f$ satisfiable?
\end{minipage}
\end{center}

A decision problem in \NP\, class is \NP-\emph{complete} if
it is not less difficult than any other \NP\, problem.
\begin{lemma}\label{lemma:CKT-SAT-NP-C}
\cite{AB12}
The CIRCUIT SATISFIABILITY problem is \NP-complete.
\end{lemma}
A  decision problem $P$ is \emph{polynomial-time Karp reducible} to a decision problem $Q$ if
there is a polynomial-time computable transformation $T$
mapping instances of $P$ to those of $Q$
such that an instance $x$ of $P$ answers yes if and only if $T(x)$
answers yes\cite{AB12}.
A decision problem is \NP-\emph{hard} if a \NP-complete problem is polynomial-time
Karp reducible to it\cite{AB12}.

An FSR is completely characterized by its feedback logic.
We use Boolean circuits to characterize the
feedback logic of FSRs for
the following two reasons\footnote{
Some theorists adopt the term \textquotedblleft propositional directed acyclic graph(PDAG)\textquotedblright, 
and a PDAG is essentially the same as a Boolean circuit.}.
First, FSRs are 
mostly implemented with silicon chips, and
the Boolean circuit is an abstract 
model of their feedback logic in silicon chips.
Second,
the Boolean circuit is a generalization of Boolean formula\cite{AB12}.
Therefore, in this correspondence the size of an FSR is measured by
the size of its feedback logic as a Boolean circuit.

\subsection{The cycle structure of FSRs}
A binary sequence $\seq{s}$ is a map from $\Int$ to $\BnrSet$.
If there exists some $\tau\in\Int$ such that
$\seq{s}(t+\tau)=\seq{s}(t)$ for any $t\in\Int$,
$\seq{s}$ is said to be \emph{periodic}
and the \emph{period} of $\seq{s}$ is defined to be
\begin{equation*}
\per{\seq{s}} =
\min\set{\tau>0:\seq{s}(t+\tau)=\seq{s}(t) \text{ for all }t\in\Int}.
\end{equation*}

Let $f$ be an $m$-stage FSR.
The following three statements are equivalent 
\cite{Gol67}:
(i) The state transformation of $f$ is bijective.
(ii) Any sequence generated by $f$ is 
periodic.
(iii) $\crct{f}(x_0,x_1,\dots,x_m)=x_m\oplus\crct{g}(x_1,x_2,\dots,x_{m-1})\oplus x_0$
for some $(m-1)$-input Boolean function $g$.
If any of (i)-(iii) holds, $f$ is said to be \emph{nonsingular}.

In the rest of this section we only consider nonsingular FSRs.

A sequence $\seq{s}$ of period $m$ determines a cyclic sequence
$\seqcyc{\seq{s}}=[\seq{s}(0),\seq{s}(1),\dots,\seq{s}(m-1)]$.
We call $\seqcyc{\seq{s}}$ to be an  $m$-cycle
and also denote $\lengthof{\seqcyc{\seq{s}}}=m$.
For the $m$-cycle $\seqcyc{\seq{s}}$, define the set
\begin{equation*}
\cycsecset{k}{\seqcyc{\seq{s}}}=
\set{\left(\seq{s}(i),\seq{s}\left((i+1)\bmod m\right),\dots,\seq{s}\left((i+k-1)\bmod m\right)\right)\in\BnrSet^k:0\leq i<m}.
\end{equation*}
Actually, any shift of a periodic sequence determines the same cycle,
and $\set{\seq{s}':\seqcyc{\seq{s}'}=\seqcyc{\seq{s}}}$
is exactly the set of all shifts of $\seq{s}$.
Furthermore, if $\seq{s}\in\SeqSet{f}$ for a $k$-stage  FSR $f$,
then 
each vector in $\cycsecset{k}{\seqcyc{\seq{s}}}$ plays as a unique initial state
and hence determines a unique sequence in $\set{\seq{s}':\seqcyc{\seq{s}'}=\seqcyc{\seq{s}}}$.

The \emph{cycle structure} of an FSR ${f}$, denoted by $\CycleStr{f}$, is
$\set{\seqcyc{\seq{s}}:\seq{s}\in\SeqSet{{f}}}$.

Following this definition, we have the lemma below.
\begin{lemma}\label{lemma:subFSR-cyc-str}
Let $f$ and $g$ be FSRs.
Then $g$ is a subFSR of $f$ if and only if $\CycleStr{g}\subset\CycleStr{f}$.
\end{lemma}

\begin{lemma}\label{lemma:conjugate-nominkiss}
Let $f$ be an $m$-stage FSR.
Suppose $\cycle{c},\cycle{d}\in\CycleStr{f}$(including $\cycle{c}=\cycle{d}$),
$\state{u}\in\cycsecset{m}{\cycle{c}}$ and
$\vecconjugate{\state{u}}\in\cycsecset{m}{\cycle{d}}$.
Then  $\min\left(\cycsecset{m}{\cycle{c}}\cup\cycsecset{m}{\cycle{d}}\right)<\min\set{\state{u},\vecconjugate{\state{u}}}$
or $\state{u}\in\set{\vecnull{m},\veconenull{m}}$.
\end{lemma}
\pf 
Let $F$ denote the state transformation of the FSR $f$.
Then $\set{F(\state{u}),F(\vecconjugate{\state{u}})}
=\set{\relint{\state{u}/2},
\relint{\state{u}/2}+2^{m-1}}$,
where $\relint{\state{u}/2}=\max\set{i\in\Int:i\leq\state{u}/2}$.

Notice that $F(\state{u})\in\cycsecset{m}{\cycle{c}}$,
$F(\vecconjugate{\state{u}})\in\cycsecset{m}{\cycle{d}}$
and $\set{\state{u},\vecconjugate{\state{u}}}=\set{2\relint{\state{u}/2},2\relint{\state{u}/2}+1}$.
If $\relint{\state{u}/2}>0$, then
$\relint{\state{u}/2}<\min\set{\state{u},\vecconjugate{\state{u}}}$,
implying  
\begin{equation*}
\min\left(\cycsecset{m}{\cycle{c}}\cup\cycsecset{m}{\cycle{d}}\right)
\le \min\set{F(\state{u}),F(\vecconjugate{\state{u}})}  <\min\set{\state{u},\vecconjugate{\state{u}}}.
\end{equation*}
If $\relint{\state{u}/2}=0$, then $\state{u}\in\set{\vecnull{m},\veconenull{m}}$.
\fp

\begin{lemma}\label{lemma:cycle-structure-fsr}
Let $\CycleSet{C}$ be a set of cycles.
Then there exists an $m$-stage FSR ${f}$
with $\CycleStr{f}=\CycleSet{C}$
if and only if the following two conditions hold:
(i) $\sum_{\cycle{c}\in\CycleSet{C}}\lengthof{\cycle{c}} = 2^m$;
(ii) The map $\state{v}\mapsto\lbit{m}{\state{v}}$ is injective on
${\bigcup_{\cycle{c}\in\CycleSet{C}}\cycsecset{m+1}{\cycle{c}}}$.
\end{lemma}
To prove Lemma \ref{lemma:cycle-structure-fsr}, we
use the following Lemma.
\begin{lemma}\label{lemma:cycle-structure-fsr-cnd}
Let $\CycleSet{C}$ be a set of finitely many cycles.
Then the following three statements are equivalent:
(i) $\cset{\bigcup_{\cycle{c}\in\CycleSet{C}}\cycsecset{m}{\cycle{c}}}=\sum_{\cycle{c}\in\CycleSet{C}}\lengthof{\cycle{c}}$;
(ii) The map $\state{v}\mapsto\lbit{m}{\state{v}}$
is injective on
${\bigcup_{\cycle{c}\in\CycleSet{C}}\cycsecset{m+1}{\cycle{c}}}$;
(iii) The map $\state{v}\mapsto\hbit{m}{\state{v}}$ is injective on
${\bigcup_{\cycle{c}\in\CycleSet{C}}\cycsecset{m+1}{\cycle{c}}}$.
\end{lemma}
\pf First we prove that Statements (i) and (ii) are equivalent.

Let $\CycleSet{C}=\set{\cycle{c}_1,\cycle{c}_2,\dots,\cycle{c}_k}$
and $\cycle{c}_i=[c_{i,0},c_{i,1},\dots,c_{i,p_i-1}]$, $1\leq i\leq k$,
where $p_i=\lengthof{\cycle{c}_i}$.
In this proof, a tuple $(i,j)$ denotes a pair of integers
satisfying $1\leq i\leq k$ and $0\leq j<p_i$.
Denote
\begin{align*}
\state{x}_{i,j}=&(c_{i,(j+1)\bmod p_i},c_{i,(j+2)\bmod p_i},\dots,c_{i,(j+m)\bmod p_i});\\
\state{y}_{i,j}=&(c_{i,j},c_{i,(j+1)\bmod p_i},c_{i,(j+2)\bmod p_i},\dots,c_{i,(j+m)\bmod p_i}).
\end{align*}
Notice
$\bigcup_{\cycle{c}\in\CycleSet{C}}\cycsecset{m}{\cycle{c}}
=\bigcup_{i=1}^k \set{ \state{x}_{i,j}: 0\leq j< p_i}$ and 
$\bigcup_{\cycle{c}\in\CycleSet{C}}\cycsecset{m+1}{\cycle{c}}
=\bigcup_{i=1}^k \set{ \state{y}_{i,j}: 0\leq j< p_i}$.
It is sufficient to consider cases below.
\begin{itemize}
  \item  Case $\cset{\bigcup_{\cycle{c}\in\CycleSet{C}}\cycsecset{m}{\cycle{c}}}=\sum_{\cycle{c}\in\CycleSet{C}}\lengthof{\cycle{c}}$.
   Then
$\state{x}_{i,j} = \state{x}_{i',j'}$ if and only if $(i,j)=(i',j')$.
Since $\state{x}_{i,j}=\lbit{m}{\state{y}_{i,j}}$,
$\state{y}_{i,j} = \state{y}_{i',j'}$ occurs only if $(i,j)=(i',j')$.
That is, the map $\state{y}_{i,j}  \mapsto  \lbit{m}{\state{y}_{i,j}}=\state{x}_{i,j}$ is injective on
${\bigcup_{\cycle{c}\in\CycleSet{C}}\cycsecset{m+1}{\cycle{c}}}$.
  \item Case $\cset{\bigcup_{\cycle{c}\in\CycleSet{C}}\cycsecset{m}{\cycle{c}}}\neq\sum_{\cycle{c}\in\CycleSet{C}}\lengthof{\cycle{c}}$.
   Then
$\state{x}_{i_0,j_0} = \state{x}_{i_0',j_0'}$ for some $(i_0,j_0)\neq(i_0',j_0')$.

\emph{Claim:}
If $\state{x}_{i,j_0} = \state{x}_{i',j_0'}$ for some $(i,j_0)\neq(i',j_0')$,
then there exists $(i,j_1)$ and $(i',j_1')$ such that
$\state{x}_{i,j_1} = \state{x}_{i',j_1'}$
and $\state{y}_{i,j_1} \neq \state{y}_{i',j_1'}$.

\emph{Proof of the claim.}
Assume that this claim does not hold. Then
for any $(i,j_1)$ and $(i',j_1')$, if
$\state{x}_{i,j_1} = \state{x}_{i',j_1'}$ then
$\state{y}_{i,j_1} = \state{y}_{i',j_1'}$.
Notice that $\state{y}_{i,j} = \state{y}_{i',j'}$ implies
$\state{x}_{i,(j-1)\bmod p_i} = \state{x}_{i',(j'-1)\bmod p_{i'}}$.
Then
$\state{x}_{i,(j_0-t)\bmod p_i} = \state{x}_{i',(j_0'-t)\bmod p_{i'}}$
 for any $t\geq0$.
Hence, 
$\cycle{c}_i=\cycle{c}_{i'}$
 and  $p_i\mid (j_0'-j_0)$,
 contradictory to $(i,j_0)\neq(i',j_0')$.
Therefore, our assumption is absurd and the claim is proved.

Following this claim, we assume
$\state{x}_{i_0,j_0} = \state{x}_{i_0',j_0'}$ and
$\state{y}_{i_0,j_0} \neq \state{y}_{i_0',j_0'}$ for some
$(i_0,j_0)\neq(i_0',j_0')$.
Thus,  the map $\state{v}  \mapsto  \lbit{m}{\state{v}}$ is not injective on
${\bigcup_{\cycle{c}\in\CycleSet{C}}\cycsecset{m+1}{\cycle{c}}}$.
\end{itemize}

The proof of equivalence of Statements (i) and (iii)
is similar and we omit it here.
\fp
\pf[Proof of Lemma \ref{lemma:cycle-structure-fsr}.] By Lemma \ref{lemma:cycle-structure-fsr-cnd},
it is sufficient to prove this statement:
$\CycleStr{f}=\CycleSet{C}$
if and only if
$\cset{\bigcup_{\cycle{c}\in\CycleSet{C}}\cycsecset{m}{\cycle{c}}}=
\sum_{\cycle{c}\in\CycleSet{C}}\lengthof{\cycle{c}}=
2^m$.

Suppose $\CycleSet{C}=\CycleStr{{f}}$ for some $m$-stage FSR ${f}$.
Then for any $\cycle{c}\in\CycleSet{C}$,
a vector in
$\cycsecset{m}{\cycle{c}}$
is exactly an initial state and uniquely determines a sequence in $\SeqSet{f}$.
Thus, $\bigcup_{\cycle{c}\in\CycleSet{C}}\cycsecset{m}{\cycle{c}}=\BnrSet^k$
and $\cset{\bigcup_{\cycle{c}\in\CycleSet{C}}\cycsecset{m}{\cycle{c}}}=
\sum_{\cycle{c}\in\CycleSet{C}}\lengthof{\cycle{c}}$.

Suppose  $\cset{\bigcup_{\cycle{c}\in\CycleSet{C}}\cycsecset{m}{\cycle{c}}}=
\sum_{\cycle{c}\in\CycleSet{C}}\lengthof{\cycle{c}}=
2^m$.
Then
$\bigcup_{\cycle{c}\in\CycleSet{C}}
{\cycsecset{m}{\cycle{c}}}=\BnrSet^{m}$.
Define an $m$-input Boolean function $f_1$ as follows.
By Lemma \ref{lemma:cycle-structure-fsr-cnd},
for any $\state{v}=(a_0,a_1,\dots,a_{m-1})\in\BnrSet^m$,
there exists uniquely $b\in\BnrSet$
such that $(a_0,a_1,\dots,a_{m-1},b)\in
\bigcup_{\cycle{c}\in\CycleSet{C}}\cycsecset{m+1}{\cycle{c}}$.
We define $f_1(\state{v})=b$.
Immediately, $\CycleSet{C}$ is the cycle structure of an FSR
whose feedback logic is logically equivalent to $f_1$.
\fp

\begin{lemma}\label{lemma:cycle-state-cycle}
Let $f$ be an $m$-stage FSR and $F$ the state transformation of $f$.
Let $\cycle{c}\in\CycleStr{f}$ and $\per{\cycle{c}}=p$.
Then for any $\state{v}\in\cycsecset{m}{\cycle{c}}$,
$\min\set{i>0:F^i(\state{v})=\state{v}}=p$
and $\cycsecset{m}{\cycle{c}}=\set{\state{v},
F(\state{v}),\dots,F^{p-1}(\state{v})}$.
\end{lemma}
\pf
Let $\state{v}\in\cycsecset{m}{\cycle{c}}$
and $q=\min\set{i>0:F^i(\state{v})=\state{v}}$.
Clearly, $q\leq p$.
Then
\begin{equation*}
{\cycle{c}}=[\hbit{1}{\state{v}},
\hbit{1}{F(\state{v})},\dots,\hbit{1}{F^{q-1}(\state{v})}],
\end{equation*}
and
$q=\per{\cycle{c}}=p$.
Because $\set{F^i(\state{v}):i\in\Int}\subseteq \cycsecset{m}{\cycle{c}}$
and $\cset{\cycsecset{m}{\cycle{c}}}\leq \per{\cycle{c}}$,
we conclude that $\cset{\cycsecset{m}{\cycle{c}}}=p$
and $\cycsecset{m}{\cycle{c}}=\set{\state{v},
F(\state{v}),\dots,F^{p-1}(\state{v})}$ is a set of $p$ vectors in $\BnrSet^m$.
\fp

\begin{lemma}\label{lemma:cycle-join}
Let $g(x_0,x_1,\dots,x_m)$ be an $m$-stage FSR and
\begin{equation*}
f(x_0,x_1,\dots,x_m)=g(x_0,x_1,\dots,x_m)\oplus f_3(x_1,x_2,\dots,x_{m-1}),
\end{equation*}
where $f_3$ is an $(m-1)$-input Boolean logic.
Let $\lambda:\BnrSet^m\rightarrow\BnrSet$ be a map satisfying
\begin{equation}\label{eqn:cycle-join-lambda-cnd}
\left\{
\begin{aligned}
&\cset{\set{{\state{v}}\in\cycsecset{m}{\cycle{c}}:\cycextr{\state{v}}=1}}\leq1
\text{ for any }\cycle{c}\in\CycleStr{g};\\
 & 
{  \cycextr{\state{v}}\cdot\cycextr{\vecconjugate{\state{v}}}=0
 \text{ for any }\state{v}\in\BnrSet^m;}\\
& \text{For any }\state{u}\in\BnrSet^{m-1}
\text{ with }
f_3({\state{u}})=1, \text{ there exists }b\in\BnrSet
\text{ satisfying }\cycextr{b\parallel\state{u}}=1.
\end{aligned}
\right.
\end{equation}
A directed 
graph $D_g^f$ is defined as follows:
the set of vertices is $\CycleStr{g}$,
and an arc is incident from $\cycle{c}_1$ to $\cycle{c}_2$ if and only if
\begin{equation*}
{\set{\state{v}\in\cycsecset{m}{\cycle{c}_1}:
f_3(\lbit{m-1}{\state{v}})=1,\cycextr{\state{v}}=1,
\vecconjugate{\state{v}}\in \cycsecset{m}{\cycle{c}_2}}}\neq\emptyset.
\end{equation*}
If $D_g^f$ is acyclic,
then the following two statements hold:
(i)
Any $\cycle{d}\in\CycleStr{f}$
is joined by all cycles in a weakly connected component\footnote{
Let $D$ be a directed graph  with its set of vertices $V$.
An undirected graph $H$ is obtained by taking each arc of $D$
as an edge of $H$. The weakly connected component(s) is(are)
the connected component(s) of $H$. Formally,
define a binary relation
\begin{equation*}
R=\set{(a,b)\in V\times V:\text{ there is an arc incident from }
a \text{ to }b \text{ or there is an arc incident from }
b \text{ to }a},
\end{equation*}
and then a weakly connected component of
$D$ is an equivalence class w.r.t.
the equivalence closure of $R$.}
$\CycleSet{C}$ of $D_g^f$
and
$\cycsecset{m}{\cycle{d}}=\bigcup_{\cycle{c}\in\CycleSet{C}
}\cycsecset{m}{\cycle{c}}$.
(ii) 
If $h$ is a subFSR of $f$, then
$\CycleStr{h}\subset\CycleStr{g}$.
\end{lemma}
\pf
Statement (i) of this lemma follows from the idea of the cycle joining method\cite{Gol67},
and we leave its proof in Appendix \ref{appnd:proof-cycle-join}.
Below we prove Statement (ii) of this lemma.

By Lemmas \ref{lemma:subFSR-cyc-str} and \ref{lemma:cycle-structure-fsr},
it is sufficient to prove this statement:
if $\CycleSet{C}\subset\CycleStr{f}$
and $\CycleSet{C}\not\subset\CycleStr{g}$,
then for any $1\leq k< m$, the map $\state{v}\mapsto\lbit{k}{\state{v}}$ is not injective on
$\bigcup_{\cycle{c}\in\CycleSet{C}}\cycsecset{k+1}{\cycle{c}}$.
Suppose  $\cycle{d}\in\CycleSet{C}
\setminus\CycleStr{g}$.
As proved in Statement (i),
$\cycle{d}$ is joined by the cycles
composing a weakly connected component $\CycleSet{D}$ of the graph $D_{g}^f$.
Since $\CycleSet{D}\subset\CycleStr{g}$
and $\cycle{d}\notin\CycleStr{g}$, we have
$\cset{\CycleSet{D}}>1$.
Hence, by Statement (i) and the definition of $D_{g}^{f}$, 
there exists $\state{v}\in\BnrSet^{m}$ satisfying
$\set{\state{v},\vecconjugate{\state{v}}}\subset\cycsecset{m}{\cycle{d}}$.
Then for any $1\le k<m$, 
$\hbit{k+1}{\state{v}},\hbit{k+1}{\vecconjugate{\state{v}}}
\in\cycsecset{k+1}{\cycle{d}}$
satisfy
$\hbit{k+1}{\state{v}}\neq\hbit{k+1}{\vecconjugate{\state{v}}}$
and
$\lbit{k}{\hbit{k+1}{\state{v}}}=
 \lbit{k}{\hbit{k+1}{\vecconjugate{\state{v}}}}$.
Therefore, the map $\state{v}\mapsto\lbit{k}{\state{v}}$ is not injective on
$\cycsecset{k+1}{\cycle{d}}$,
and hence is not injective on $\bigcup_{\cycle{c}\in\CycleSet{C}}\cycsecset{k+1}{\cycle{c}}$.
\fp

Given an $m$-cycle $\cycle{c}=[b_0,b_1,\dots,b_m]$,
let  $\veccomplement{\cycle{c}}$ denote the cycle
$[b_0\oplus1,b_1\oplus1,\dots,b_m\oplus1]$.

The cycle structure of LFSRs is well understood.
\begin{lemma}
\label{lemma:poly-irreducible}
\label{lemma:cycle-structure-lfsr-p0}
\label{lemma:cycle-structure-lfsr-p0x}
Let $\len=3^{k}$, $0\leq k\in\Int$.
Let $p_0(x)=x^{2\len} \oplus x^{\len} \oplus1$,
$p_1(x)=(x\oplus1)\cdot p_0(x)$,
and $p_2(x)=x^{4\len}\oplus x^{2\len}\oplus 1$
be polynomials over the binary field $\GF{2}$.
Then $p_0$
is irreducible over $\GF{2}$ and 
\begin{align*}
\CycleStr{p_0}=&\set{[0],
\cycle{\beta}_1,\cycle{\beta}_2,\dots,\cycle{\beta}_{\frac{2^{2\len}-1}{3\len}}},\\
\CycleStr{p_1}=&\set{[0],\cycle{\beta}_1,\cycle{\beta}_2,\dots,\cycle{\beta}_{\frac{2^{2\len}-1}{3\len}},
[1],\veccomplement{\cycle{\beta}_1},\veccomplement{\cycle{\beta}_2},\dots,\veccomplement{\cycle{\beta}_{\frac{2^{2\len}-1}{3\len}}}},\\
\CycleStr{p_2}=&\set{[0],
\cycle{\beta}_1,\cycle{\beta}_2,\dots,\cycle{\beta}_{\frac{2^{2\len}-1}{3\len}},
\cycle{\gamma}_1,\cycle{\gamma}_2,\dots,\cycle{\gamma}_{\frac{2^{4\len}-2^{2\len}}{6\len}}},
\end{align*}
where
$\per{\cycle{\beta}_i}=\per{\veccomplement{\cycle{\beta}_i}}=3\len$ for $1\leq i\leq \frac{2^{2\len}-1}{3\len}$,
and $\per{\cycle{\gamma}_i}=6\len$ for $1\leq i\leq \frac{2^{4\len}-2^{2\len}}{6\len}$.
\end{lemma}
\pf Since $p_0(x)\cdot(x^{3^k}\oplus1)=
x^{3^{k+1}}\oplus 1$ and $\gcd(p_0,x^{3^k}\oplus1)=1$,
the roots of $p_0$ are exactly primitive $3^{k+1}$-th roots of unity.
Thus, $p_0$ is irreducible and $\min\set{0<t\in\Int:p_0\mid(x^t-1)}=3\len$
is the order of any primitive $3^{k+1}$-th root of unity
in the multiplicative group of the finite field $\GF{2}[x]/(p_0(x))$.

The rest of this lemma directly follows from 
\cite[Theorem 8.53, 8.55, 8.63]{LN83}.
\fp

\section{\NP-hardness of deciding irreducible FSRs}\label{sect:FSR-irreducibility}
Below Algorithm \ref{alg:fsr-red}
transforms a given Boolean circuit to an FSR.

\begin{algorithm}
\caption{Transforming a Boolean circuit to an FSR}
\label{alg:fsr-red}
\renewcommand{\algorithmicrequire}{\textbf{Input:}}
\renewcommand{\algorithmicensure}{\textbf{Output:}}
\begin{algorithmic}[1]
\REQUIRE An $r$-input Boolean circuit $\crct{f_0}$.
\ENSURE\label{line:len-red} A $4\len$-stage FSR $f$,
where $k=\min\set{i\in\Int:i\geq\log_3(r/2)}$
and $\len=3^{k}$.
\STATE
\COMMENT{%
Construct a $(4\len-1)$-input Boolean circuit $\crct{f_3}$
with its pseudocode in Lines \ref{line:C1-begin}-\ref{line:C1-end}.
In the rest of this section,
$L$ denotes the state transformation of the LFSR
               ${x^{4\len}\oplus x^{2\len}\oplus 1}$.
}
\STATE\label{line:C1-begin}
Let $\state{x}
\in\BnrSet^{4\len-1}$
be the input of $\crct{f_3}$.
\STATE Let
$\state{u}_0=0\parallel\state{x}
$ and
$\state{v}_0=1\parallel\state{x}
$.
\FOR{$i=1$ to $6\len$}
    \STATE $\state{u}_i=L(\state{u}_{i-1})$ and
               $\state{v}_i=L(\state{v}_{i-1})$.
    \STATE
        $a_i={f_0}(\lbit{r}{\state{u}_i})$
        and
        $b_i={f_0}(\lbit{r}{\state{v}_i})$.
    \IF{$L^{3\len}(\vecconjugate{\state{u}_i})= \vecconjugate{\state{u}_i}$ or
    $L^{6\len}(\vecconjugate{\state{u}_i})\neq\min\set{L^j(\vecconjugate{\state{u}_i}):1\leq j\leq 6\len}$}
        \STATE $c_i=1$.
    \ELSE
        \STATE $c_i=0$.
    \ENDIF
    \IF{$L^{3\len}(\vecconjugate{\state{v}_i})= \vecconjugate{\state{v}_i}$ or
    $L^{6\len}(\vecconjugate{\state{v}_i})\neq\min\set{L^j(\vecconjugate{\state{v}_i}):1\leq j\leq 6\len}$}
        \STATE $d_i=1$.
    \ELSE
        \STATE $d_i=0$.
    \ENDIF
\ENDFOR
\STATE\label{line:max}
        $\state{u}_{\min}=\min\set{{\state{u}}_i:1\leq i\leq 6\len}$ and
        $\state{v}_{\min}=\min\set{{\state{v}}_i:1\leq i\leq 6\len}$.
\IF{$c_1\myAND c_2\myAND \cdots\myAND c_{6\len}=1$
and ${\state{u}_\len}= \state{u}_{\min}$
and $L^{3\len}(\vecconjugate{\state{u}_{\min}})=\vecconjugate{\state{u}_{\min}}$
}
    \STATE $q(\state{u}_0)=1$.
\ELSE
    \STATE $q(\state{u}_0)=0$.
\ENDIF
\IF{$d_1\myAND d_2\myAND \cdots\myAND d_{6\len}=1$
and ${\state{v}_\len}= \state{v}_{\min}$
and $L^{3\len}(\vecconjugate{\state{v}_{\min}})=\vecconjugate{\state{v}_{\min}}$
}
    \STATE $q(\state{v}_0)=1$.
\ELSE
    \STATE $q(\state{v}_0)=0$.
\ENDIF
\IF{$\state{u}_0 = \state{u}_{3\len} $ and $ \state{u}_{6\len}=\state{u}_{\min}$ and $
a_1\myOR a_2\myOR \cdots\myOR a_{6\len}=1$}
    \STATE The Boolean circuit $\crct{f_3}$ returns $1$.
\ELSIF{$\state{v}_0 = \state{v}_{3\len}$ and $ \state{v}_{6\len} = \state{v}_{\min}$ and $
b_1\myOR b_2\myOR \cdots\myOR b_{6\len}=1$}
    \STATE The Boolean circuit $\crct{f_3}$ returns $1$.
\ELSIF{$\state{u}_0 \neq \state{u}_{3\len} $ and $\state{v}_0 \neq \state{v}_{3\len} $
      and ($ \state{u}_{6\len}=\state{u}_{\min}$ or $ \state{v}_{6\len} = \state{v}_{\min}$
      or $q(\state{u}_0)=1$ or $q(\state{v}_0)=1$)}
    \STATE The Boolean circuit $\crct{f_3}$ returns $1$.
\ELSE
    \STATE The Boolean circuit $\crct{f_3}$ returns $0$.
\ENDIF\label{line:C1-end}
    \RETURN
     the FSR $f(x_0,\dots,x_{4\len})=x_{4\len}\oplus x_{2\len}\oplus x_0
     \oplus \crct{f_3}(x_1,x_2,\dots,x_{4\len-1})$.
\end{algorithmic}
\end{algorithm}

In the rest of this section, we use notations
$\crct{f_0}$, $\crct{f_3}$ and $f$ 
defined in Algorithm \ref{alg:fsr-red}.

Clearly, $f$ is a nonsingular FSR.

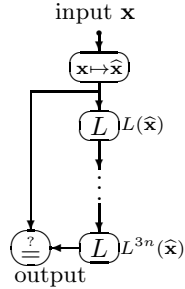
\begin{figure}[htbp]
\begin{center}
\setlength{\unitlength}{1.3mm}
\begin{picture}(17,29)
\put(0,10){\usebox{\myCPcircuit}}
\end{picture}
\end{center}
\caption{A diagram of the subcircuit CP}\label{fig:subcircuitCP}
\end{figure}

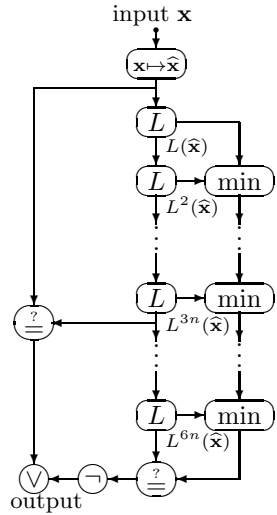
\begin{figure}[htbp]
\begin{center}
\setlength{\unitlength}{1.3mm}
\begin{picture}(26,53)
\put(0,21){\usebox{\myCMPcircuit}}
\end{picture}
\end{center}
\caption{A diagram of the subcircuit CMP}\label{fig:subcircuitCMP}
\end{figure}

Following Algorithm \ref{alg:fsr-red},
the Boolean circuit $f_3$ 
is described with
Figures \ref{fig:subcircuitCP}, \ref{fig:subcircuitCMP},
\ref{fig:subcircuitMQ}, \ref{fig:subcircuitPS}
and \ref{fig:circuit-f3}.
To ease our presentation, from now on
we also use operations
with finite fan-in and fan-out
for sketching a Boolean circuit.
For example, as $x\oplus y=((\myNOT x)\myAND y)\myOR((\myNOT y)\myAND x)$,
we allow XOR($\myXOR$),
logically equivalent to a subcircuit consisting of
 five gates.
In Figures \ref{fig:subcircuitCP}, \ref{fig:subcircuitCMP},
\ref{fig:subcircuitMQ} and \ref{fig:subcircuitPS},
the operation \textquotedblleft$\IsEqual{?}$\textquotedblright\, decides whether two $4\len$-bit inputs are equal or not.
In Figures \ref{fig:subcircuitCMP} and \ref{fig:subcircuitMQ},
the operation \textquotedblleft$\min$\textquotedblright\, computes the minimum of two $4\len$-bit integers.
\begin{figure}[htbp]
\begin{center}
\setlength{\unitlength}{1.3mm}
\begin{picture}(32,58) 
\put(0,30){\usebox{\myMQcircuit}}
\put(0,2){\makebox(0,0)[l]{\scriptsize Input: $\state{x}\in\BnrSet^{4\len}$}}
\put(0,-1){\makebox(0,0)[l]{\scriptsize Output: $m(\state{x}),q(\state{x})\in\BnrSet$}}
\put(16,-4){\makebox(0,0){\scriptsize The subcircuits
CP and CMP are given in Figures \ref{fig:subcircuitCP}
and \ref{fig:subcircuitCMP}, respectively.}}
\end{picture}
\end{center}
\caption{A diagram of the subcircuit MQ}\label{fig:subcircuitMQ}
\end{figure}
\begin{figure}[htbp]
\begin{center}
\setlength{\unitlength}{1.3mm}
\begin{picture}(28,46)
\put(0,23){\usebox{\myPScircuit}}
\put(0,-1){\makebox(0,0)[l]{\scriptsize Input: $\state{x}\in\BnrSet^{4\len}$}}
\put(0,-4){\makebox(0,0)[l]{\scriptsize Output: $p(\state{x}),s(\state{x})\in\BnrSet$}}
\end{picture}
\end{center}
\caption{A diagram of the subcircuit PS}\label{fig:subcircuitPS}
\end{figure}
\begin{figure}[htbp]
\begin{center}
\setlength{\unitlength}{1.3mm}
\begin{picture}(44,28)
\put(22,10.5){\usebox{\myREDcircuit}}
\put(22,-4){\makebox(0,0){\scriptsize The subcircuits
MQ and PS are given in Figures \ref{fig:subcircuitMQ} and \ref{fig:subcircuitPS}, respectively.}}
\end{picture}
\end{center}
\caption{A diagram of the Boolean circuit $\crct{f_3}$}\label{fig:circuit-f3}
\end{figure}

\begin{lemma}\label{lemma:alg-red-poly-time}
Let $f_1$ be the feedback logic of the FSR $f$
given by Algorithm \ref{alg:fsr-red}.
Then $\sizeof{{f_1}}< 37908\cdot\sizeof{f_0}^4$
and Algorithm \ref{alg:fsr-red} is polynomial-time computable.
\end{lemma}
\pf
The operation $\state{x}\mapsto\vecconjugate{\state{x}}$ uses 
one NOT gate on $\hbit{1}{\state{x}}$.
Given the input  $(x_0,x_1,\dots,x_{4\len-1})$
and $(y_0,y_1,\dots,y_{4\len-1})$,
the operation \textquotedblleft$\IsEqual{?}$\textquotedblright\
outputs
$\myNOT((x_0\oplus y_0)\myOR(x_1\oplus y_1)\myOR\dots\myOR(x_{4\len-1}\oplus y_{4\len-1}))$
and costs at most $24\len$ gates.
The state transformation $L$
is performed by one XOR gate, i.e., $5$ gates.
By Appendix \ref{appnd:min},
the operation \textquotedblleft$\min$\textquotedblright\,
uses $104\len^2 + 66\len - 22$ gates.

Noticing  $r\leq 2\len\leq 3r-1$, $r\leq \sizeof{\crct{f_0}}$ and
\begin{equation*}
f_1(x_0,\dots,x_{4\len-1})=x_0\oplus x_{2\len}\oplus \crct{f_3}(x_1,x_2,\dots,x_{4\len-1}),
\end{equation*}
we count gates in Figure \ref{fig:circuit-f3} and obtain
\begin{align*}
\sizeof{\crct{f_1}}=&11+\sizeof{\crct{f_3}}\\
=&12\len\cdot\sizeof{f_0} + 7488\len^4 + 4752\len^3 - 856\len^2 + 274\len + 69\\
<& 37908\cdot\sizeof{f_0}^4.
\end{align*}

The Boolean circuit $f_0$
has $\sizeof{f_0}$ vertices and
less than $2\cdot\sizeof{f_0}$ arcs;
The feedback logic $f_1$ has
at most $37908\cdot\sizeof{f_0}^4$ vertices
and at most $75816\cdot\sizeof{f_0}^4$ arcs.
The FSR $f$ uses $f_0$ and
basic polynomial-time computable operations for
at most $37908\cdot\sizeof{f_0}^4$ times
and its main architecture is given by Figures \ref{fig:subcircuitCP},
\ref{fig:subcircuitCMP},
\ref{fig:subcircuitMQ},
\ref{fig:subcircuitPS}
and \ref{fig:circuit-f3}.
Therefore,
Algorithm \ref{alg:fsr-red}
is polynomial-time computable.
\fp

In the rest of this section,
$\len$ is as given in Algorithm \ref{alg:fsr-red},
$p_0$ and $p_2$ are
the polynomials as defined in Lemma \ref{lemma:poly-irreducible},
we also denote
$\CycleSet{C}_{6\len}=\CycleStr{p_2}\setminus\CycleStr{p_0}$.

\begin{lemma}\label{lemma:ell-conjugate-2ell}
Let $\state{v}\in\cycsecset{4\len}{\cycle{\beta}}$,
where $\cycle{\beta}\in\CycleStr{p_0}$.
Then $\vecconjugate{\state{v}}\in\cycsecset{4\len}{\cycle{\gamma}}$
for some $\cycle{\gamma}\in\CycleSet{C}_{6\len}$.
\end{lemma}
\pf Suppose $\vecconjugate{\state{v}}\in\cycsecset{4\len}{\cycle{\gamma}}$
for some $\cycle{\gamma}\in\CycleStr{p_0}$.
By Lemmas \ref{lemma:cycle-state-cycle} and \ref{lemma:cycle-structure-lfsr-p0},
$L^{3\len}(\vecconjugate{\state{v}})=\vecconjugate{\state{v}}$
and
$L^{3\len}(\state{v})=\state{v}$.
Since $L$ is a linear transformation and $\veconenull{4\len}=\state{v}\oplus
\vecconjugate{\state{v}}$,
we have
$L^{3\len}(\veconenull{4\len})=\veconenull{4\len}$,
contradictory to
$L^{3\len}(\veconenull{4\len})=
(\vecnull{\len}1\vecnull{2\len-1}1\vecnull{\len-1})$,
where this vector is written without commas between bits.
Therefore, the supposition above is absurd
and $\cycle{\gamma}\in\CycleSet{C}_{6\len}$.
\fp

Because any $\state{v}\in\BnrSet^{4\len}$, as an initial state of
the $4\len$-stage LFSR $p_2$, determines a unique cycle,
in the rest of this section we denote
$\veccyc{\state{v}}=\cycle{c}\in\CycleStr{p_2}$ such that
$\state{v}\in\cycsecset{4\len}{\cycle{c}}$.
\begin{lemma}\label{lemma:cycle-structure-fsr-red-f-Vset}
Let
\begin{align*}
\CycleSet{D}=&\left\{
\cycle{c}\in\CycleStr{p_2}:
\veccyc{
\min\cycsecset{4\len}{\cycle{c}}\oplus\veconenull{4\len}
}\in\CycleStr{p_0},
\right.\\
&\qquad\qquad\qquad\qquad\left.\left\{\state{v}\in\cycsecset{4\len}{\cycle{c}}:
\veccyc{\vecconjugate{\state{v}}}\in\CycleSet{C}_{6\len}
\text{ and }
\vecconjugate{\state{v}}
=\min\cycsecset{4\len}{\veccyc{\vecconjugate{\state{v}}}}\right\}=\emptyset\right\}
\end{align*}
and define a map
$\rho:\CycleStr{p_2}\rightarrow\BnrSet^{4\len}$
as
\begin{equation*}\label{eq:1}
\cycvec{\cycle{c}}=
\left\{
\begin{array}{l@{\text{ if }}l}
\min \cycsecset{4\len}{\cycle{c}},& \cycle{c}\in\CycleStr{p_2}\setminus\CycleSet{D};\\
L^{5\len}\left(\min\cycsecset{4\len}{\cycle{c}}\right),&\cycle{c}\in\CycleSet{D}.
\end{array}
\right.
\end{equation*}
Then the following two statements hold:
(i) $\CycleSet{D}\subset\CycleSet{C}_{6\len}$ and
for any $\cycle{c}\in\CycleSet{D}$,
 $\veccyc{\vecconjugate{\cycvec{\cycle{c}}}}\in \CycleSet{C}_{6\len}\setminus\CycleSet{D}$.
(ii) If  $\cycle{c}\in\CycleStr{p_2}$ and $\vecconjugate{\cycvec{\cycle{c}}}\in
\set{\cycvec{\cycle{e}}:\cycle{e}\in\CycleStr{p_2}}$,
then $\cycle{c}\in\set{[0],\veccyc{\veconenull{4\len}}}$.
\end{lemma}
\pf
For convenience in this proof we may write a cycle or vector
without commas between its bits.\\
\emph{Claim:} If $\cycle{c}\in\CycleStr{p_2}$
satisfies $\veccyc{\min\cycsecset{4\len}{\cycle{c}}\oplus\veconenull{4\len}
}\in\CycleStr{p_0}$,
 then
   $\cycle{c}=[1\state{u}_{0}
   0\state{u}_{1}0\state{u}_{2}
   0\state{u}_{0}1\state{u}_{1}
   0\state{u}_{2}]$, where
   $\state{u}_0,\state{u}_1,\state{u}_2\in\BnrSet^{\len-1}$,
   $\state{u}_2=\state{u}_0\oplus \state{u}_1$
   and
   $(1\state{u}_{0}
   0\state{u}_{1}0\state{u}_{2}
   0\state{u}_{0})=\min\cycsecset{4\len}{\cycle{c}}$.\\
\emph{Proof of the claim.}
By Lemma \ref{lemma:ell-conjugate-2ell}, $\cycle{c}\in\CycleSet{C}_{6\len}$.
Denote
$\cycle{c}=[a_0\state{u}_0a_1\state{u}_1
a_2\state{u}_2a_3\state{u}_3
a_4\state{u}_4a_5\state{u}_5]$,
where
\begin{equation*}
\left\{
\begin{aligned}
& a_i\in\BnrSet,0\leq i\leq 5;\\
&\state{u}_i\in\BnrSet^{\len-1}, 0\leq i\leq 5;\\
&(a_0\state{u}_0a_1\state{u}_1
a_2\state{u}_2a_3\state{u}_3)=\min\cycsecset{4\len}{\cycle{c}}.
\end{aligned}
\right.
\end{equation*}
Notice $\veccyc{\veconenull{4\len}}=[1\vecnull{4\len-1}1\vecnull{2\len-1}]$.
Then
$\veccomplement{a_0}\state{u}_0a_1\state{u}_1
a_2\state{u}_2a_3\state{u}_3
\veccomplement{a_4}\state{u}_4a_5\state{u}_5$
is concatenation of a same cycle in $\CycleStr{p_0}$, implying
where $a_2=a_0\oplus a_1\oplus1$ and $\state{u}_2=\state{u}_0\oplus \state{u}_1$.
By Lemma \ref{lemma:cycle-structure-lfsr-p0},
\begin{equation*}
    \cycle{c}=[a_0\state{u}_0a_1\state{u}_1a_2\state{u}_2
\veccomplement{a_0}\state{u}_0\veccomplement{a_1}\state{u}_1a_2\state{u}_2].
\end{equation*}
By
\begin{equation*}
(a_0\state{u}_0a_1\state{u}_1
a_2\state{u}_2\veccomplement{a_0}\state{u}_0)\leq
(\veccomplement{a_0}\state{u}_0\veccomplement{a_1}\state{u}_1a_2\state{u}_2
a_0\state{u}_0),
\end{equation*}
we have $a_0=1$.
By
\begin{equation*}
(a_0\state{u}_0a_1\state{u}_1
a_2\state{u}_2\veccomplement{a_0}\state{u}_0)\leq
(\state{u}_0a_1\state{u}_1
a_2\state{u}_2\veccomplement{a_0}\state{u}_0\veccomplement{a_1}),
\end{equation*}
we have $a_1=0$. Then $a_2=0$.
The proof of this claim is complete.

For a $k\len$-cycle  $\cycle{c}=[b_0,b_1,\dots,b_{k\len-1}]$,
we call
\begin{equation*}
(b_{i},b_{(i+\len)\bmod k\len},b_{(i+2\len)\bmod k\len},\dots,b_{(i+(k-1)\len)})
\end{equation*}
an \emph{$\len$-sampling} of $\cycle{c}$,
$0\leq i< k\len$.

Choose any $\cycle{c}\in\CycleSet{D}$.
Because of the claim above,
let   $\cycle{c}=[1\state{u}_{0}
   0\state{u}_{1}0\state{u}_{2}
   0\state{u}_{0}1\state{u}_{1}
   0\state{u}_{2}]$, where
   $\state{u}_0,\state{u}_1,\state{u}_2\in\BnrSet^{\len-1}$,
   $\state{u}_2=\state{u}_0\oplus \state{u}_1$
    and
   $(1\state{u}_{0}
   0\state{u}_{1}0\state{u}_{2}
   0\state{u}_{0})=\min\cycsecset{4\len}{\cycle{c}}$.
Then
\begin{equation*}
\vecconjugate{\cycvec{\cycle{c}}}=\veconenull{4\len}\oplus L^{5\len}\left(\min\cycsecset{4\len}{\cycle{c}}\right)=
(1\state{u}_{2}1\state{u}_{0}
   0\state{u}_{1}0\state{u}_{2})
\end{equation*}
   and hence
\begin{equation*}
\veccyc{\vecconjugate{\cycvec{\cycle{c}}}}=
[
1\state{u}_{2}
   1\state{u}_{0}
   0\state{u}_{1}0\state{u}_{2}
   1\state{u}_{0}1\state{u}_{1}
].
\end{equation*}
First, $3\len\nmid\lengthof{\veccyc{\vecconjugate{\cycvec{\cycle{c}}}}}$.
By Lemma \ref{lemma:cycle-structure-lfsr-p0},
$\veccyc{\vecconjugate{\cycvec{\cycle{c}}}}\in \CycleSet{C}_{6\len}$.
Second, as shown in the claim above,
there is an $\len$-sampling $(100010)$ of any cycle in $\CycleSet{D}$,
while $(100010)$ is not an $\len$-sampling of
$\veccyc{\vecconjugate{\cycvec{\cycle{c}}}}$.
Hence, $\veccyc{\vecconjugate{\cycvec{\cycle{c}}}}\notin \CycleSet{D}$.
By Lemma \ref{lemma:ell-conjugate-2ell}, $\cycle{c}\notin\CycleStr{p_0}$.
Till now Statement (i) of this lemma is proved.

Now we prove Statement (ii) of this lemma.

Denote $\state{v}_0=(01\vecnull{4\len-2})$.
Then $\veccyc{\vecconjugate{\state{v}_0}}
=[11\vecnull{4\len-2}11\vecnull{2\len-2}]$.
By Lemma \ref{lemma:cycle-structure-lfsr-p0},
$\veccyc{\vecconjugate{\state{v}_0}}\in\CycleSet{C}_{6\len}$.
Seeing $\vecconjugate{\state{v}_0}=\min
\cycsecset{4\len}{\veccyc{
\vecconjugate{\state{v}_0}}
}$
and  $\state{v}_0\in\cycsecset{4\len}{\veccyc{\veconenull{4\len}}}$,
we have 
$\veccyc{\veconenull{4\len}}
\notin\CycleSet{D}$
and
$\veconenull{4\len}=
\cycvec{\veccyc{\veconenull{4\len}}}$.

Denote $V_c=\set{\cycvec{\cycle{e}}:\cycle{e}\in\CycleStr{p_2}}$.
Notice that $\cycvec{\cycle{c}}\in\cycsecset{4\len}{\cycle{c}}$,
$\cycle{c}\in\CycleStr{p_2}$.
It is sufficient to consider the following cases.
\begin{enumerate}
  \item
\label{case:cycextr-1} If $\cycle{c}\in\CycleSet{D}$,
by Statement (i), $\veccyc{\vecconjugate{\cycvec{\cycle{c}}}}\in\CycleSet{C}_{6\len}\setminus\CycleSet{D}$.
By the definition of $\CycleSet{D}$,
\begin{equation*}
\vecconjugate{\cycvec{\cycle{c}}}\neq \min\cycsecset{4\len}{\veccyc{\vecconjugate{\cycvec{\cycle{c}}}}}
=\cycvec{\veccyc{\vecconjugate{\cycvec{\cycle{c}}}}}
\end{equation*}
and hence $ \vecconjugate{\cycvec{\cycle{c}}}\notin V_c$.
  \item
\label{case:cycextr-2} If $\veccyc{\veconenull{4\len}}\neq\cycle{c}\in\CycleSet{C}_{6\len}\setminus\CycleSet{D}$,
then $\cycvec{\cycle{c}}=\min\cycsecset{4\len}{\cycle{c}}\notin
\set{\vecnull{4\len},\veconenull{4\len}}$.
By the definition of $\CycleSet{D}$,
$\veccyc{\vecconjugate{\cycvec{\cycle{c}}}}\notin
\CycleSet{D}$, yielding
$\cycvec{\veccyc{\vecconjugate{\cycvec{\cycle{c}}}}}=\min\cycsecset{4\len}{\veccyc{\vecconjugate{\cycvec{\cycle{c}}}}}$.
By Lemma \ref{lemma:conjugate-nominkiss},
$\vecconjugate{\cycvec{\cycle{c}}}\neq
\min\cycsecset{4\len}{\veccyc{\vecconjugate{\cycvec{\cycle{c}}}}}$
 and hence
$\vecconjugate{\cycvec{\cycle{c}}}\notin V_c$.
  \item
\label{case:cycextr-3} If $[0]\neq\cycle{c}\in\CycleStr{p_0}$
and $\veccyc{\vecconjugate{\cycvec{\cycle{c}}}}\notin
\CycleSet{D}$,
then similar to Case (\ref{case:cycextr-2}),
we also get $\vecconjugate{\cycvec{\cycle{c}}}\notin V_c$.
  \item
\label{case:cycextr-4} If $\cycle{c}\in\CycleStr{p_0}$
and $\veccyc{\vecconjugate{\cycvec{\cycle{c}}}}\in
\CycleSet{D}$,
then by the proved Statement (i) of this lemma,
\begin{equation*}
\cycle{c}\neq\veccyc{\cycvec{\veccyc{\vecconjugate{\cycvec{\cycle{c}}}}}\oplus\veconenull{4\len}}\in\CycleSet{C}_{6\len}.
\end{equation*}
Because $\veccyc{\cycvec{\cycle{e}}}=\cycle{e}$ for
any $\cycle{e}\in\CycleStr{p_2}$, 
we have
$\vecconjugate{\cycvec{\cycle{c}}}\neq \cycvec{\veccyc{\vecconjugate{\cycvec{\cycle{c}}}}}$,
yielding
$\vecconjugate{\cycvec{\cycle{c}}}\notin V_c$.
 \item
\label{case:cycextr-5} Besides, consider $\cycle{c}\in\set{[0],\veccyc{\veconenull{4\len}}}$.
We have $\vecconjugate{\cycvec{[0]}}=\veconenull{4\len}=
\cycvec{\veccyc{\veconenull{4\len}}}$ since $\veccyc{\veconenull{4\len}}\in\CycleSet{C}_{6\len}\setminus\CycleSet{D}$.
\end{enumerate}
Till now all cases are listed and
Statement (ii) of this lemma holds.
\fp

\begin{lemma}\label{lemma:cycle-structure-fsr-red-f}
Let $\rho$ be given in Lemma \ref{lemma:cycle-structure-fsr-red-f-Vset}.
Let the map $\lambda:\BnrSet^{4\len}\rightarrow\BnrSet$ be defined as
\begin{equation*}
\cycextr{\state{v}}=
\left\{
\begin{aligned}
1,& \text{ if }\state{v}\in\set{\cycvec{\cycle{c}}:\cycle{c}\in\CycleStr{p_2}}
\text{ and }\veccyc{\vecconjugate{\state{v}}}\in\CycleSet{C}_{6\len};\\
0,& \text{ otherwise.}
\end{aligned}
\right.
\end{equation*}
Let $D_{p_2}^{f}$ be the graph defined
as in Lemma \ref{lemma:cycle-join}(Recall that $f$ and $f_3$
are given in Algorithm \ref{alg:fsr-red}).
Then the following statements hold:
(i) 
Statements (i) and (ii) of Lemma \ref{lemma:cycle-join}
hold, where $g$ in Lemma \ref{lemma:cycle-join}
is the LFSR $p_2$.
(ii)
Each $\cycle{c}\in\CycleSet{C}_{6\len}$ is not
an isolated vertex in $D_{p_2}^{f}$.
(iii) Every $\cycle{c}\in\CycleStr{p_0}$
is an isolated vertex in $D_{p_2}^{f}$  if and only if
$f_0$ is unsatisfiable.
\end{lemma}
\pf
Since $\cycvec{\cycle{c}}\in\cycsecset{4\len}{\cycle{c}}$
for any $\cycle{c}\in \CycleStr{p_2}$, we have
\begin{equation*}
\cset{\set{\state{v}\in\cycsecset{4\len}{\cycle{c}}:\cycextr{\state{v}}=1}}
\le 
\cset{\set{\cycvec{\cycle{c}}}}=1.
\end{equation*}

Following from Statement (ii) of Lemma \ref{lemma:cycle-structure-fsr-red-f-Vset}
and $\cycextr{\veconenull{4\len}}=0$,
we have
$\cycextr{\state{v}}\cdot \cycextr{\vecconjugate{\state{v}}}=0$
for any $\state{v}\in\BnrSet^{4\len}$.

Use $\CycleSet{D}$ defined in Lemma \ref{lemma:cycle-structure-fsr-red-f-Vset}.

Let $q(\state{u}_0)$, $q(\state{v}_0)$,
$\state{u}_i$ and $\state{v}_i$ be as in Algorithm \ref{alg:fsr-red},
$0\leq i\leq 6\len$.
Denote $\cycle{e}_0=\veccyc{\state{u}_0}$
and $\cycle{e}_1=\veccyc{\state{v}_0}$.
By Lemmas \ref{lemma:cycle-state-cycle} and \ref{lemma:cycle-structure-lfsr-p0},
$\cycsecset{4\len}{\cycle{e}_0}=\set{\state{u}_i:1\leq i\leq 6\len}$;
$\cycsecset{4\len}{\cycle{e}_1}=\set{\state{v}_i:1\leq i\leq 6\len}$;
$\state{u}_0=\state{u}_{6\len}$;
$\state{v}_0=\state{v}_{6\len}$;
$\state{u}_0=\state{u}_{3\len}$(resp. $\state{v}_0=\state{v}_{3\len}$)
if and only if $\cycle{e}_0\in\CycleStr{p_0}$(resp. $\cycle{e}_1\in\CycleStr{p_0}$);
$L^{3\len}(\vecconjugate{\state{u}_i})=\vecconjugate{\state{u}_i}$(resp. $L^{3\len}(\vecconjugate{\state{v}_i})=\vecconjugate{\state{v}_i}$)
if and only if $\veccyc{\vecconjugate{\state{u}_i}}
\in\CycleStr{p_0}$(resp. $\veccyc{\vecconjugate{\state{v}_i}}
\in\CycleStr{p_0}$);
$L^{6\len}(\vecconjugate{\state{u}_i})\neq \min\set{L^j(\vecconjugate{\state{u}_i}):1\leq j\leq 6\len}$(resp.
$L^{6\len}(\vecconjugate{\state{v}_i})\neq \min\set{L^j(\vecconjugate{\state{v}_i}):1\leq j\leq 6\len}$)
if and only if $\vecconjugate{\state{u}_i}\neq \min\cycsecset{4\len}{\veccyc{\vecconjugate{\state{u}_i}}}$(resp.
$\vecconjugate{\state{v}_i}\neq \min\cycsecset{4\len}{\veccyc{\vecconjugate{\state{v}_i}}}$);
$\state{u}_\len=\state{u}_{\min}$(resp. $\state{v}_\len=\state{v}_{\min}$)
if and only if $\state{u}_0=L^{5\len}\left(\min\cycsecset{4\len}{\cycle{e}_0}\right)$
(resp. $\state{v}_0=L^{5\len}\left(\min\cycsecset{4\len}{\cycle{e}_1}\right)$);
$L^{3\len}(\vecconjugate{\state{u}_{\min}})=\vecconjugate{\state{u}_{\min}}$(resp. $L^{3\len}(\vecconjugate{\state{v}_{\min}})=\vecconjugate{\state{v}_{\min}}$) is equivalent to 
$\veccyc{\vecconjugate{\state{u}_{\min}}}\in\CycleStr{p_0}$(resp. 
$\veccyc{\vecconjugate{\state{v}_{\min}}}\in\CycleStr{p_0}$).
Then $q(\state{u}_0)=1$(resp. $q(\state{v}_0)=1$) if and only if
$\cycle{e}_0\in\CycleSet{D}$ and $\state{u}_0=\cycvec{\cycle{e}_0}$%
(resp. $\cycle{e}_1\in\CycleSet{D}$ and $\state{v}_0=\cycvec{\cycle{e}_1}$).
By Lemma \ref{lemma:ell-conjugate-2ell},
$\set{\cycle{e}_0,\cycle{e}_1}\not\subset\CycleStr{p_0}$.
Then $f_3(\lbit{4\len-1}{\state{u}_0})=f_3(\lbit{4\len-1}{\state{v}_0})=1$
if and only if one of the following cases holds:
\begin{enumerate}
  \item\label{case:1} $\cycle{e}_0\in\CycleStr{p_0}$,
$\state{u}_0 = \min\cycsecset{4\len}{\cycle{e}_0}$
and 
$\set{\state{v}\in\cycsecset{4\len}{\cycle{e}_0}: f_0(\lbit{r}{\state{v}})=1}\neq\emptyset$;
  \item\label{case:2} $\cycle{e}_1\in\CycleStr{p_0}$,
$\state{v}_0 = \min\cycsecset{4\len}{\cycle{e}_1}$
and $\set{\state{v}\in\cycsecset{4\len}{\cycle{e}_1}: f_0(\lbit{r}{\state{v}})=1}\neq\emptyset$;
 \item\label{case:3} $\cycle{e}_0\in\CycleSet{C}_{6\len}$,
 $\state{u}_0 = \min\cycsecset{4\len}{\cycle{e}_0}$
 and $\veccyc{\vecconjugate{\state{u}_0}}=\cycle{e}_1\in\CycleSet{C}_{6\len}$;
 \item\label{case:4} $\cycle{e}_1\in\CycleSet{C}_{6\len}$,
 $\state{v}_0 = \min\cycsecset{4\len}{\cycle{e}_1}$
 and $\veccyc{\vecconjugate{\state{v}_0}}=\cycle{e}_0\in\CycleSet{C}_{6\len}$;
 \item\label{case:5} $\cycle{e}_0,\cycle{e}_1\in\CycleSet{C}_{6\len}$,
$\cycle{e}_0\in\CycleSet{D}$ and $\state{u}_0=\cycvec{\cycle{e}_0}$;
 \item\label{case:6} $\cycle{e}_0,\cycle{e}_1\in\CycleSet{C}_{6\len}$,
$\cycle{e}_1\in\CycleSet{D}$ and $\state{v}_0=\cycvec{\cycle{e}_1}$.
\end{enumerate}
Considering Statement (i) of Lemma \ref{lemma:cycle-structure-fsr-red-f-Vset},
we have
\begin{equation}\label{eqn:red-f3-equiv}
    f_3(\state{x})=\left\{
    \begin{array}{lll}
    1,&\text{ if } \state{x}=\lbit{4\len-1}{\state{v}}, \text{ where }
    \state{v}=\cycvec{\cycle{c}} \text{ and } \cycle{c},
    \veccyc{\vecconjugate{\state{v}}}\in\CycleSet{C}_{6\len};
&\text{ by Cases \ref{case:3}, \ref{case:4}, \ref{case:5} and \ref{case:6}}\\
    1,&\text{ if } \state{x}=\lbit{4\len-1}{\state{v}}, \text{ where }
    \state{v}=\cycvec{\cycle{c}}, \cycle{c}\in\CycleStr{p_0}\\
    &\text{ and }
    \set{\state{u}\in\cycsecset{4\len}{\cycle{c}}:f_0(\lbit{r}{\state{u}})=1}\neq\emptyset;
&\text{ by Cases \ref{case:1} and \ref{case:2}}\\
    0,&\text{ otherwise.}
&    \end{array}
    \right.
\end{equation}
By Eq.(\ref{eqn:red-f3-equiv}) and 
Lemma \ref{lemma:ell-conjugate-2ell},
$ f_3(\state{x})=1$ implies that there exists $\state{v}\in\BnrSet^{4\len}$
satisfying $\state{x}=\lbit{4\len-1}{\state{v}}$
and $\cycextr{\state{v}}=1$.

Hitherto we have shown that Eq.(\ref{eqn:cycle-join-lambda-cnd}) holds,
where $g$ in Eq.(\ref{eqn:cycle-join-lambda-cnd}) is the LFSR $p_2$.

By Lemma \ref{lemma:conjugate-nominkiss} and
Statement (i) of Lemma \ref{lemma:cycle-structure-fsr-red-f-Vset},
$D_{p_2}^f$ is loopless.
Assume that $D_{p_2}^f$ is not acyclic.
Then in $D_{p_2}^f$ there is a walk
$(\cycle{c}_0,\cycle{c}_1,\dots,\cycle{c}_{m-1},\cycle{c}_{m})$
for some $m\ge2$.
Specifically, $\cycle{c}_i\in\CycleStr{p_2}$, $0\leq i<m$, are pairwise distinct,
$\cycle{c}_{m}=\cycle{c}_0$,
and for any $0\leq i< m$
there is an arc 
incident from $\cycle{c}_{i}$ to $\cycle{c}_{i+1}$.
By Lemma \ref{lemma:ell-conjugate-2ell} and the definition of $\lambda$,
any $\cycle{c}\in\CycleStr{p_0}$ is a source in
$D_{p_2}^f$.
Additionally,
for $\cycle{c}\in\CycleSet{D}$,
by Statement (i) of Lemma \ref{lemma:cycle-structure-fsr-red-f-Vset}
and 
\begin{equation*}
\left\{\state{v}\in\cycsecset{4\len}{\cycle{c}}:
\veccyc{\vecconjugate{\state{v}}}\in\CycleSet{C}_{6\len}
\text{ and }
\vecconjugate{\state{v}}
=\min\cycsecset{4\len}{\veccyc{\vecconjugate{\state{v}}}}\right\}=\emptyset,
\end{equation*}
no arc is incident from a cycle in $\CycleSet{C}_{6\len}$
to $\cycle{c}$, i.e.,
any arc entering $\cycle{c}$
leaves from a source in $\CycleStr{p_0}$.
Besides, as shown in the proof of Lemma \ref{lemma:cycle-structure-fsr-red-f-Vset}, $\cycvec{\veccyc{\veconenull{4\len}}}=\veconenull{4\len}$
and $\veccyc{\vecconjugate{\veconenull{4\len}}}=[0]$,
then $\veccyc{\veconenull{4\len}}$ is a sink in $D_{p_2}^f$.
Therefore, $\veccyc{\veconenull{4\len}}\neq\cycle{c}_i\in\CycleSet{C}_{6\len}\setminus \CycleSet{D}$
and $\cycvec{\cycle{c}_i}=\min\cycsecset{4\len}{\cycle{c}_i}$,
 $0\leq i<m$.
Noticing $\vecconjugate{\cycvec{\cycle{c}_i}}\in
\cycsecset{4\len}{\cycle{c}_{i+1}}$, $0\leq i<m$,
by Lemma \ref{lemma:conjugate-nominkiss}, we have
$\min\cycsecset{4\len}{\cycle{c}_{i+1}}<\min\cycsecset{4\len}{\cycle{c}_{i}}$
for $0\leq i< m$, implying $\min\cycsecset{4\len}{\cycle{c}_0}<
\min\cycsecset{4\len}{\cycle{c}_0}$,
which is ridiculous.
Therefore, the assumption is absurd
and $D_{p_2}^f$ is acyclic.

Till now 
we have proved that Eq.(\ref{eqn:cycle-join-lambda-cnd}) holds and $D_{p_2}^{f}$ is acyclic, where $g$ is the LFSR $p_2$.
By Lemma \ref{lemma:cycle-join},
Statement (i) of this lemma is proved.

Now we prove Statement (ii) of this lemma.
Suppose $\cycle{c}\in\CycleSet{C}_{6\len}$.
\begin{itemize}
  \item
If $\veccyc{\veconenull{4\len}\oplus\min\cycsecset{4\len}{\cycle{c}}}\in
\CycleSet{C}_{6\len}$, then $\cycle{c}\notin\CycleSet{D}$,
$\cycvec{\cycle{c}}=\min\cycsecset{4\len}{\cycle{c}}$
and $\cycextr{\cycvec{\cycle{c}}}=1$. By Eq.(\ref{eqn:red-f3-equiv}),
an arc leaves $\cycle{c}$.
 \item
 If $\veccyc{\veconenull{4\len}\oplus\min\cycsecset{4\len}{\cycle{c}}}\in
\CycleStr{p_0}$ and
\begin{equation*}
\left\{\state{v}\in\cycsecset{4\len}{\cycle{c}}:
\veccyc{\vecconjugate{\state{v}}}\in\CycleSet{C}_{6\len}
\text{ and }
\vecconjugate{\state{v}}
=\min\cycsecset{4\len}{\veccyc{\vecconjugate{\state{v}}}}\right\}\neq\emptyset.
\end{equation*}
Let $\state{v}_0\in\cycsecset{4\len}{\cycle{c}}$ satisfy
$\veccyc{\vecconjugate{\state{v}_0}}\in\CycleSet{C}_{6\len}$
and $\vecconjugate{\state{v}_0}
=\min\cycsecset{4\len}{\veccyc{\vecconjugate{\state{v}_0}}}$.
Clearly, $\veccyc{\vecconjugate{\state{v}_0}}\notin\CycleSet{D}$.
Then
$\cycvec{\veccyc{\vecconjugate{\state{v}_0}}}
=\min\cycsecset{4\len}{\veccyc{\vecconjugate{\state{v}_0}}}
=\vecconjugate{\state{v}_0}$
and $\cycextr{\vecconjugate{\state{v}_0}}=1$. By Eq.(\ref{eqn:red-f3-equiv}),
an arc enters $\cycle{c}$.
 \item
If $\veccyc{\veconenull{4\len}\oplus\min\cycsecset{4\len}{\cycle{c}}}\in
\CycleStr{p_0}$ and
\begin{equation*}
\left\{\state{v}\in\cycsecset{4\len}{\cycle{c}}:
\veccyc{\vecconjugate{\state{v}}}\in\CycleSet{C}_{6\len}
\text{ and }
\vecconjugate{\state{v}}
=\min\cycsecset{4\len}{\veccyc{\vecconjugate{\state{v}}}}\right\}=\emptyset,
\end{equation*}
then $\cycle{c}\in\CycleSet{D}$.
By Statement (i) of Lemma \ref{lemma:cycle-structure-fsr-red-f-Vset}
and Eq.(\ref{eqn:red-f3-equiv}),
an arc is incident from  $\cycle{c}$ to
a cycle in $\CycleSet{C}_{6\len}\setminus\CycleSet{D}$.
\end{itemize}
Till now Statement (ii) of this lemma is proved.

Now we prove Statement (iii) of this lemma.
Since $\cycextr{\state{v}}=1$ occurs only if
$\veccyc{\vecconjugate{\state{v}}}\in\CycleSet{C}_{6\len}$,
in $D_{p_2}^f$ no arc enters any $\cycle{c}\in\CycleStr{p_0}$.
Since $r\leq2\len$ and
$\bigcup_{\cycle{c}\in\CycleStr{p_0}}\cycsecset{2\len}{\cycle{c}}=\BnrSet^{2\len}$,
we have
\begin{equation*}
\set{\lbit{r}{\state{v}}:\state{v}\in\bigcup_{\cycle{c}\in\CycleStr{p_0}}\cycsecset{4\len}{\cycle{c}}}=
\set{\lbit{r}{\state{v}}:\state{v}\in\bigcup_{\cycle{c}\in\CycleStr{p_0}}\cycsecset{2\len}{\cycle{c}}}
=\BnrSet^r.
\end{equation*}
By Lemma \ref{lemma:ell-conjugate-2ell}, 
$\cycextr{\cycvec{\cycle{c}}}=1$
for any $\cycle{c}\in\CycleStr{p_0}$.
Then by Eq.(\ref{eqn:red-f3-equiv}),
in $D_{P_2}^f$ there exists an arc incident from  some $\cycle{c}\in\CycleStr{p_0}$
to some $\cycle{d}\in\CycleSet{C}_{6\len}$
if and only if $f_0$ is satisfiable.
Thus, Statement (iii) of this lemma is proved.
\fp

\begin{lemma}\label{lemma:subFSR-order}
If $g$ is a subFSR of the FSR $f$, then $g$ is of stage $2\len$.
\end{lemma}
\pf Let $g$ be of stage $m$.
By Lemmas \ref{lemma:cycle-structure-fsr},
$\sum_{\cycle{d}\in\CycleStr{g}}\per{\cycle{d}}=2^m$.
By Lemmas \ref{lemma:subFSR-cyc-str}, \ref{lemma:cycle-structure-lfsr-p0}
 and Statement (i) of \ref{lemma:cycle-structure-fsr-red-f},
for any $\cycle{d}\in\CycleStr{g}$, 
$\per{\cycle{d}}\equiv\cset{\set{\vecnull{4\len}}\cap\cycsecset{4\len}{\cycle{d}}}\bmod 3\len$.
Then we have an integer equation $a+3\len b=2^m$,
where $a\in\BnrSet$ and $0\leq b<2^{4\len}/(3\len)$.
Since $2\len=\min\set{i>0:{3\len}\mid(2^i-1)}$, where $\len=3^k$ for some $1\leq k\in\Int$,
we have $a=1$ and $2\len\mid m$. Hence, $m=2\len<4\len$.
\fp

\begin{lemma}\label{lemma:reduction-irreducible}
The FSR $f$ is irreducible
if and only if the Boolean circuit $f_0$ is satisfiable.
\end{lemma}
\pf
Suppose $f_0$ to be unsatisfiable.
By 
Statements (i) and (iii) of Lemma \ref{lemma:cycle-structure-fsr-red-f},
$\CycleStr{p_0}\subset\CycleStr{f}$.
By Lemma \ref{lemma:subFSR-cyc-str},
$p_0$ is a subFSR of $f$ and hence $f$ is reducible.

Suppose $f_0$ to be satisfiable.
Assume that $h$ is a subFSR of $f$.
By Statement (i) of Lemma \ref{lemma:cycle-structure-fsr-red-f},
\begin{equation}\label{eqn:red-irr-1}
\CycleStr{h}\subset\CycleStr{p_2}.
\end{equation} 
Furthermore,
by Statements (i) and (ii) of Lemma \ref{lemma:cycle-structure-fsr-red-f}, 
any cycle in $\CycleSet{C}_{6\len}$ joins with other cycles to combine a cycle in $\CycleStr{f}$,
implying
\begin{equation}\label{eqn:red-irr-2}
\left(\CycleStr{p_2}\setminus\CycleStr{p_0}\right)\cap \CycleStr{f}=\emptyset.
\end{equation} 
Similarly, by Statements (i) and (iii) of Lemma \ref{lemma:cycle-structure-fsr-red-f},
if $f_0$ is satisfiable, then
\begin{equation}\label{eqn:red-irr-3}
\CycleStr{p_0}\not\subset\CycleStr{f}.
\end{equation}
By Eqs.(\ref{eqn:red-irr-1}), (\ref{eqn:red-irr-2}), (\ref{eqn:red-irr-3}) and Lemma \ref{lemma:subFSR-cyc-str}, we get
\begin{equation*}
\CycleStr{h}\subset\CycleStr{f}\cap\CycleStr{p_2}
\subset\CycleStr{f}\cap\CycleStr{p_0}
\subsetneq\CycleStr{p_0}.
\end{equation*}
By Lemma \ref{lemma:subFSR-order}, $h$ is of stage $2\len$.
However, by Lemma \ref{lemma:cycle-structure-fsr},
\begin{equation*}
2^{2\len}=\sum_{\cycle{c}\in\CycleStr{h}}\per{\cycle{c}}<
\sum_{\cycle{c}\in\CycleStr{p_0}}\per{\cycle{c}}=2^{2\len},
\end{equation*}
which is absurd.
Therefore,  $f$ is irreducible.
\fp

\begin{center}
\begin{minipage}[h]{0.7\textwidth}
\textbf{PROBLEM}:  FSR IRREDUCIBILITY

{INSTANCE}: An FSR $f$ with its feedback logic $f_1$ as a Boolean circuit of size $\sizeof{f_1}$.

{QUESTION}: Is $f$ irreducible?
\end{minipage}
\end{center}

By Lemmas \ref{lemma:CKT-SAT-NP-C}, 
\ref{lemma:alg-red-poly-time}
and \ref{lemma:reduction-irreducible},
Algorithm \ref{alg:fsr-red} is a polynomial-time Karp reduction
from 
CIRCUIT SATISFIABILITY
to 
FSR IRREDUCIBILITY.
Therefore, we conclude that
\begin{theorem}\label{thm:irreducibility-np-hard}
The 
FSR IRREDUCIBILITY
problem is \NP-hard.
\end{theorem}

\section{ \NP-hardness of deciding indecomposable FSRs
}\label{sect:FSR-indecomposability}

\begin{lemma}\label{lemma:f2-f0-satisfiability}
Let $f_0$ be an $r$-input Boolean logic and
\begin{equation}\label{eqn:f0-f2}
f_2(\state{x})=\left\{
\begin{array}{ll}
  0, & \text{ if } \state{x}=\vecnull{r}; \\
  1, & \text{ if } \state{x}=\vecone{r} \text{ and } f_0(\vecone{r})=1; \\
  f_0(\vecnull{r}), &\text{ if }
  \state{x}=\vecone{r}  \text{ and } f_0(\vecone{r})=0; \\
  f_0(\state{x}), & \text{ otherwise. }
\end{array}
\right.
\end{equation}
Then the Boolean function
$\crct{f_2}$ is satisfiable
if and only if $\crct{f_0}$ is satisfiable.
\end{lemma}

Below Algorithm \ref{alg:fsr-dec}
transforms a given Boolean circuit to an FSR.

\begin{algorithm}
\caption{Transforming a Boolean circuit to an FSR}
\label{alg:fsr-dec}
\renewcommand{\algorithmicrequire}{\textbf{Input:}}
\renewcommand{\algorithmicensure}{\textbf{Output:}}
\begin{algorithmic}[1]
\REQUIRE An $r$-input Boolean circuit $\crct{f_0}$.
\ENSURE A $(2\len+1)$-stage FSR $f$, where $k=\min\set{i\in\Int:i\geq\log_3(r/2)}$
and $\len=3^{k}$.
\STATE
Construct an $r$-input Boolean circuit $\crct{f_2}$ defined by Eq.(\ref{eqn:f0-f2}).
\STATE
\COMMENT{%
Construct a $2\len$-input Boolean circuit $\crct{f_3}$
with its  pseudocode in Lines \ref{line:Cdec-begin}-\ref{line:Cdec-end}.
In the rest of this section,
$L$ denotes the state transformation of the LFSR
               ${x^{2\len}\oplus x^{\len}\oplus 1}$.
}
\STATE\label{line:Cdec-begin}
Let $(x_{1},x_{2},\dots,x_{2\len})$ be the input of $\crct{f_3}$.
\STATE\label{line:u0} $\state{u}_0=(x_{2\len}\oplus x_{\len}\oplus x_1,x_1\oplus x_2,
x_2\oplus x_3,\dots,x_{2\len-1}\oplus x_{2\len})$.
\FOR{$i=1$ to $3\len$}
    \STATE $\state{u}_i=L(\state{u}_{i-1})$.
    \STATE $a_i=\crct{f_2}(\lbit{r}{\state{u}_i})$.
\ENDFOR
    \IF{$\state{u}_{3\len} = \min\set{{{\state{u}}_i}:1\leq i\leq 3\len}$ and 
$a_1\myOR a_2\myOR \cdots\myOR a_{3\len}=1$}
       \STATE The Boolean circuit $\crct{f_3}$ returns $1$.
   \ELSE
       \STATE The Boolean circuit $\crct{f_3}$ returns $0$.
   \ENDIF\label{line:Cdec-end}
    \RETURN
     the FSR $f(x_0,\dots,x_{2\len+1})=x_{2\len+1}
     \oplus x_{2\len}\oplus x_{\len+1}\oplus x_{\len}\oplus x_1 \oplus x_0
     \oplus  \crct{f_3}(x_1,x_2,\dots,x_{2\len})$.
\end{algorithmic}
\end{algorithm}

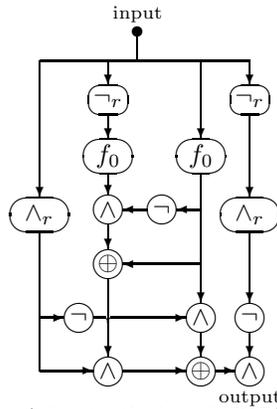
\begin{figure}[htbp]
\begin{center}
\setlength{\unitlength}{1.3mm}
\begin{picture}(26,47.5) 
\put(1,
18){\usebox{\myCnull}}
\put(13,-4){\makebox(0,0){\scriptsize Here \textquotedblleft$\myNOT_r$\textquotedblright(resp. \textquotedblleft$\myAND_r$\textquotedblright) denotes the logical NOT(resp. AND) 
 of $r$ bits.}}
\end{picture}
\end{center}
\caption{A diagram of the Boolean circuit $f_2$}\label{fig:f2}
\end{figure}

Figure \ref{fig:f2} is a sketch of $f_2$.
Following Algorithm \ref{alg:fsr-dec},
we describe $f_3$ with Figure \ref{fig:DEC-f3}.
\begin{figure}[htbp]
\begin{center}
\setlength{\unitlength}{1.3mm}
\begin{picture}(30,41)
\put(0,23){\usebox{\myDECcircuit}}

\put(-12,1){\makebox(0,0)[l]{\scriptsize 
$M$ is defined in Line \ref{line:u0} of Algorithm \ref{alg:fsr-dec};}}
\put(-12,-1){\makebox(0,0)[l]{\scriptsize
\textquotedblleft$\IsEqual{?}$\textquotedblright\, decides whether two $2\len$-bit inputs are equal or not;
}}
\put(-12,-4){\makebox(0,0)[l]{\scriptsize
\textquotedblleft$\min$\textquotedblright\, computes the minimum of two $2\len$-bit integers.
}}
\end{picture}
\end{center}
\caption{A diagram of the Boolean circuit $f_3$}\label{fig:DEC-f3}
\end{figure}

In the rest of this section, we use notations
$\crct{f_0}$, $f_2$, $\crct{f_3}$ and $f$ 
defined in Algorithm \ref{alg:fsr-dec}.

Clearly, $f$ is a nonsingular FSR.

Similar to  Lemma \ref{lemma:alg-red-poly-time}, 
we count gates in Figure \ref{fig:DEC-f3}
and derive the lemma below.
\begin{lemma}\label{lemma:alg-dec-poly-time}
Let $f_1$ be the feedback logic of the FSR $f$
given by Algorithm \ref{alg:fsr-dec}.
Then $\sizeof{{f_1}}\leq 264\cdot\sizeof{\crct{f_0}}^3$.
Particularly, Algorithm \ref{alg:fsr-dec} is 
polynomial-time computable.
\end{lemma}

In the rest of this section, 
$\len$ is given in Algorithm \ref{alg:fsr-dec},
$p_0$ and $p_1$ are
the polynomials as defined in Lemma \ref{lemma:poly-irreducible},
and we denote $\veccomplement{\CycleStr{p_0}}=\CycleStr{p_1}\setminus\CycleStr{p_0}$.
Moreover, let $L_1$ denote the state transformation of the LFSR $p_1$.

For $\state{v}=(v_0,v_1,\dots,v_{2\len})\in\BnrSet^{2\len+1}$,
define the map $\cycctrl{\state{v}}=
(v_0\oplus v_{1},v_1\oplus v_{2},\dots,v_{2\len-1}\oplus v_{2\len})\in\BnrSet^{2\len}$
and $\cycchar{\state{v}}=v_0\oplus v_{\len}\oplus v_{2\len}$.

The maps $\pi$ and  $\chi$ have the properties in Lemma \ref{lemma:conjugate-property}.
\begin{lemma}\label{lemma:conjugate-property}\label{lemma:cycle-2-family}
The following statements hold.
(i)
For 
$\state{v}\in\BnrSet^{2\len+1}$,
$\cycchar{\vecconjugate{\state{v}}}
=\cycchar{\veccomplement{\state{v}}}
=\cycchar{{\state{v}}}\oplus1$,
$\cycctrl{\vecconjugate{\state{v}}}=\vecconjugate{\cycctrl{\state{v}}}$
and $L(\cycctrl{\state{v}})=\cycctrl{L_1(\state{v})}$.
(ii)
For 
$\state{w}=(w_0,w_1,\dots,w_{2\len-1})\in\BnrSet^{2\len}$,
$\set{\state{v}\in\BnrSet^{2\len+1}:\cycctrl{\state{v}}=\state{w}}=
\set{\state{u},\veccomplement{\state{u}}}$,
where
\begin{equation*}
\state{u}=\left(0,w_0, w_0\oplus w_1,\dots,w_0\oplus w_1\oplus\cdots\oplus w_{2\len-1}\right).
\end{equation*}
(iii) For any $\state{v}\in\BnrSet^{2\len+1}$, 
\begin{equation*}
\cycchar{\state{v}}=
\left\{
\begin{aligned}
0,&\text{ if }\state{v}\in\bigcup_{\cycle{c}\in \CycleStr{p_0}}\cycsecset{2\len+1}{\cycle{c}},\\
1,&\text{ if }\state{v}\in\bigcup_{\cycle{c}\in \veccomplement{\CycleStr{p_0}}}\cycsecset{2\len+1}{\cycle{c}},
\end{aligned}
\right.
\end{equation*}
(iv) For 
$\state{v}\in\BnrSet^{2\len+1}$, if 
\begin{equation*}
{\state{v}}\in\bigcup_{\cycle{c}\in{\CycleStr{p_0}}}\cycsecset{2\len+1}{\cycle{c}},
\end{equation*}
then
\begin{equation*}
\vecconjugate{\state{v}}\in\bigcup_{\cycle{c}\in\veccomplement{\CycleStr{p_0}}}\cycsecset{2\len+1}{\cycle{c}}.
\end{equation*}
(v) For any $\state{v}\in\BnrSet^{2\len+1}$,
$\lbit{2\len}{L_1(\state{v})}=
L(\lbit{2\len}{\state{v}})\oplus \state{w}_0$,
where $\state{w}_0=(0,\dots,0,\cycchar{\state{v}})\in\BnrSet^{2\len}$.
\end{lemma}
\pf 
Statements (i) and (ii) of this lemma
can be proved by 
direct computation.

Denote $\state{v}=(v_0,v_1,\dots,v_{2\len})$.

If $\state{v}\in\bigcup_{\cycle{c}\in \CycleStr{p_0}}\cycsecset{2\len+1}{\cycle{c}}$.
clearly, $\cycchar{\state{v}}=v_0\oplus v_\len \oplus v_{2\len}=0$.
Suppose $\state{v}\in\bigcup_{\cycle{c}\in \veccomplement{\CycleStr{p_0}}}\cycsecset{2\len+1}{\cycle{c}}$.
By Lemma
 \ref{lemma:cycle-structure-lfsr-p0x},
$\veccomplement{\state{v}}\in\bigcup_{\cycle{c}\in \CycleStr{p_0}}\cycsecset{2\len+1}{\cycle{c}}$.
Then by Statement (i), 
$\cycchar{\state{v}}=1\oplus\cycchar{\veccomplement{\state{v}}}=1$.
Statement (iii) is proved.

By Lemma \ref{lemma:cycle-structure-lfsr-p0x},
\begin{equation*}
\left(\bigcup_{\cycle{c}\in{\CycleStr{p_0}}}\cycsecset{2\len+1}{\cycle{c}}\right)
\bigcup
\left(\bigcup_{\cycle{c}\in\veccomplement{\CycleStr{p_0}}}\cycsecset{2\len+1}{\cycle{c}}\right)
=\BnrSet^{2\len+1}.
\end{equation*}
Then Statement (iv) follows from  Statement (i) and (iii). 

Additionally, Statement (v) holds because
\begin{align*}
\lbit{2\len}{L_1(\state{v})}=&
(v_2,\dots,v_{2\len},v_{2\len}\oplus v_{\len+1}\oplus v_{\len}\oplus v_1 \oplus v_0)\\
=&(v_2,\dots,v_{2\len}, v_{\len+1}\oplus v_1\oplus \cycchar{\state{v}})\\
=&L((v_1,v_2,\dots,v_{2\len}))\oplus \state{w}_0\\
=&L(\lbit{2\len}{\state{v}})\oplus \state{w}_0. \hfill\qedhere
\end{align*}
\fp

\begin{lemma}\label{lemma:unique-subFSR-01-pre}
Let the map $\lambda:\BnrSet^{2\len+1}\rightarrow\BnrSet$ be defined as
\begin{equation*}
\cycextr{\state{v}}=
\left\{
\begin{aligned}
1,&\text{ if } \cycchar{\state{v}}=0 \text{ and }\cycctrl{\state{v}}=
\min\set{L^i(\cycctrl{\state{v}}): 1\leq i\leq 3\len};\\
0,&\text{ otherwise.}
\end{aligned}
\right.
\end{equation*}
Let $D_{p_1}^{f}$ be the graph defined
as in Lemma \ref{lemma:cycle-join}(Recall that $f$ and $f_3$
are given in Algorithm \ref{alg:fsr-dec}).
Then the following statements hold:
(i) Statements (i) and (ii) of Lemma \ref{lemma:cycle-join}
hold, where $g$ in Lemma \ref{lemma:cycle-join} is the LFSR $p_1$.
(ii) If $f_2$ is satisfiable, then $\CycleStr{p_0}\not\subset\CycleStr{f}$
and 
there exists $\state{v}\in\BnrSet^{2\len+1}$
satisfying $f_3(\lbit{2\len}{\state{v}})=1$
and $\cycchar{\state{v}}=0$.
\end{lemma}
\pf
Suppose $\state{v}\in\cycsecset{2\len+1}{\cycle{c}}$,
where $\cycle{c}\in\CycleStr{p_1}$.
By Statement (i) of Lemma \ref{lemma:conjugate-property},
Lemmas \ref{lemma:cycle-state-cycle} and \ref{lemma:cycle-structure-lfsr-p0x},
we get
\begin{equation}\label{eqn:dec-cycle-min}
\set{L^i(\cycctrl{\state{v}}):1\leq i\leq 3\len}=
\set{\cycctrl{L_1^i(\state{v})}:1\leq i\leq 3\len}=
\set{\cycctrl{\state{u}}:\state{u}\in\cycsecset{2\len+1}{\cycle{c}}}.
\end{equation}
Besides,
by Statements (ii) of Lemma \ref{lemma:conjugate-property},
there exists a unique vector $\state{u}$ in $\cycsecset{2\len+1}{\cycle{c}}$ satisfying 
$\cycctrl{\state{u}}=\min\set{\cycctrl{\state{u}}:\state{u}\in\cycsecset{2\len+1}{\cycle{c}}}$.
Thus, by Statement (iii) of Lemma \ref{lemma:cycle-2-family}, we have
\begin{equation}\label{eqn:dec-cycle-extr}
\cset{\set{\state{v}\in\cycsecset{2\len+1}{\cycle{c}}:\cycextr{\state{v}}=1}}=
\left\{
\begin{aligned}
1,&\text{ if }\state{c}\in\CycleStr{p_0};\\
0,&\text{ if }\state{c}\in\veccomplement{\CycleStr{p_0}}.
\end{aligned}
\right.
\end{equation}

By Statement (i) of Lemma \ref{lemma:conjugate-property},
$\cycextr{\state{v}}\cdot\cycextr{\vecconjugate{\state{v}}}=0$
for any $\state{v}\in\BnrSet^{2\len+1}$.

In Algorithm \ref{alg:fsr-dec},
$\state{x}=(x_1,x_2,\dots,x_{2\len})$
and $\state{u}_0=\cycctrl{\state{y}}$,
where $\state{y}=(x_{2\len}\oplus x_{\len},x_1,x_2,\dots,x_{2\len})
$ is the unique vector in $\BnrSet^{2\len+1}$
satisfying $\cycchar{\state{y}}=0$ and $\lbit{2\len}{\state{y}}=\state{x}$.
Let $\cycle{c}$ be the cycle satisfying $\state{y}\in\cycsecset{2\len+1}{\cycle{c}}$.
By Lemmas \ref{lemma:cycle-state-cycle}, 
 \ref{lemma:cycle-structure-lfsr-p0x} and
Eq.(\ref{eqn:dec-cycle-min}),
$\state{u}_{3\len}=
\min\set{L^i(\state{u}_0):1\leq i\leq 3\len}$
is equivalent to 
$\state{u}_0=\min\set{\cycctrl{\state{v}}:\state{v}\in\cycsecset{2\len+1}{\cycle{c}}}$.
By Eq.(\ref{eqn:dec-cycle-min}),
\begin{equation*}
\set{1\leq i\leq 3\len:f_2(\lbit{r}{L^i(\state{u}_0)})=1}\neq\emptyset
\end{equation*}
is equivalent to 
\begin{equation*}
\set{\state{u}\in\cycsecset{2\len+1}{\cycle{c}}:f_2(\lbit{r}{\cycctrl{\state{u}}})=1}\neq\emptyset.
\end{equation*}
Thus, by Algorithm \ref{alg:fsr-dec}, we have the following claim.\\
\emph{Claim.} $f_3(\state{x})=1$
if and only if $\cycextr{\state{y}}=1$  and 
$\set{\state{v}\in\cycsecset{2\len+1}{\cycle{c}}:f_2(\lbit{r}{\cycctrl{\state{v}}})=1}\neq\emptyset$.

If $f_3(\state{x})=1$, then $\cycextr{\state{y}}=1$ and $\lbit{2\len}{\state{y}}=\state{x}$.
Therefore, Eq.(\ref{eqn:cycle-join-lambda-cnd}) holds,
where $g$ in Lemma \ref{lemma:cycle-join} 
is the LFSR $p_1$.

Furthermore, by Statement (iv) of Lemma \ref{lemma:conjugate-property} and Eq.(\ref{eqn:dec-cycle-extr}),
any arc of $D_{p_1}^f$  is incident from a cycle in $\CycleStr{p_0}$ to a cycle in $\veccomplement{\CycleStr{p_0}}$.
Hence, $D_{p_1}^f$ is acyclic. 

Till now 
we have proved that Eq.(\ref{eqn:cycle-join-lambda-cnd}) holds and $D_{p_1}^{f}$ is acyclic, where $g$ in Eq.(\ref{eqn:cycle-join-lambda-cnd}) 
is the LFSR $p_1$.
By Lemma \ref{lemma:cycle-join},
Statements (i) and (ii) of Lemma \ref{lemma:cycle-join}
hold and
Statement (i) of this lemma is proved, where $g$ in Lemma \ref{lemma:cycle-join} is the LFSR $p_1$.

Now we prove Statement (ii) of this lemma.
Suppoe that $f_2$ is satisfiable. 
Since $D_{p_1}^f$ is acyclic, by Statement (i) of Lemma \ref{lemma:cycle-join}, it is sufficient 
to prove that not every $\cycle{c}\in\CycleStr{p_0}$
is isolated in $D_{p_1}^f$.

Following from Eq.(\ref{eqn:dec-cycle-extr}) and the claim above, for $\cycle{c}\in\CycleStr{p_0}$,
there exists $\state{v}\in\cycsecset{2\len+1}{\cycle{c}}$ 
satisfying $\cycextr{\state{v}}=1$ and $f_3(\lbit{2\len}{\state{v}})=1$
if and only if 
$\set{\state{v}\in\cycsecset{2\len+1}{\cycle{c}}:f_2(\lbit{r}{\cycctrl{\state{v}}})=1}\neq\emptyset$.

By Lemma \ref{lemma:cycle-structure-lfsr-p0x} and
 Statement (ii) of Lemma \ref{lemma:conjugate-property},
the map $\pi$ gives a bijection from $\bigcup_{\cycle{c}\in\CycleStr{p_0}}\cycsecset{2\len+1}{\cycle{c}}$
to $\BnrSet^{2\len}$.
Thus, seeing $r\leq 2\len$, we get
\begin{equation*}
\set{\lbit{r}{\cycctrl{\state{v}}}:\state{v}\in\bigcup_{\cycle{c}\in\CycleStr{p_0}} \cycsecset{2\len+1}{\cycle{c}}}
=\set{\lbit{r}{\state{v}}:\state{v}\in\BnrSet^{2\len}}
=\BnrSet^r.
\end{equation*}
Therefore, 
on one hand, there exists $\state{v}\in\BnrSet^{2\len+1}$
satisfying $f_3(\lbit{2\len}{\state{v}})=1$ and $\cycchar{\state{v}}=0$;
On the other hand,
in $D_{p_1}^f$ there exists at least one arc 
incident from  a cycle in $\CycleStr{p_0}$,
i.e.,
some $\cycle{c}\in\CycleStr{p_0}$
is not isolated in $D_{p_1}^f$.
By Statement (i) of this lemma,
$\cycle{c}$ joins with other cycles in $\CycleStr{p_1}$ to combine 
one cycle in $\CycleStr{f}$, and hence 
$\cycle{c}\notin\CycleStr{f}$, yielding 
$\CycleStr{p_0}\not\subset\CycleStr{f}$.
\fp

\begin{lemma}\label{lemma:unique-subFSR-01}
If $f_2$ is satisfiable and $g$ is a 
subFSR of $f$
satisfying 
$[0]\in\CycleStr{g}$, then 
$g$ is the LFSR $x_1\oplus x_0$, i.e.,
$\CycleStr{g}=\set{[0],[1]}$.
\end{lemma}
\pf Let $g$ be an $m$-stage subFSR of $f$.

By Lemma \ref{lemma:cycle-structure-fsr},
we have 
$2^m=\sum_{\cycle{d}\in\CycleStr{g}}\lengthof{\cycle{d}}$.
Furthermore, by Lemma  \ref{lemma:cycle-structure-lfsr-p0x}
and Statement (i) of \ref{lemma:unique-subFSR-01-pre},
we have
\begin{equation*}\label{eqn:dec-cyc-cnt}
\per{\cycle{d}}\equiv
\cset{\set{\vecnull{2\len+1},\vecone{2\len+1}}\cap\cycsecset{2\len+1}{\cycle{d}}}
\bmod 3\len.
\end{equation*}
Since for any $\state{v}\in\BnrSet^{2\len+1}$,
there exists a unique cycle $\cycle{c}\in\CycleStr{f}$
satisfying $\state{v}\in\cycsecset{2\len+1}{\cycle{c}}$,
we get an integer equation
\begin{equation}\label{eqn:order-cnt-dec}
3\len a+b=2^m,
\end{equation}
where  $1\leq m\leq2\len$,
$0\leq a\leq 2(2^{2\len}-1)/(3\len)$ and $b\in\set{0,1,2}$.
Since $2\len=\min\set{0<i\in\Int:3\len\mid(2^i-1)}$,
where $\len=3^k$ for some $1\leq k\in\Int$,
Eq.(\ref{eqn:order-cnt-dec})
holds only if (i) $b=1$ and $m=2\len$
or (ii) $b=2$ and $m=1$.
So, we only have to consider two possible cases below.

Case (i): $g$ is of stage $2\len$.
By Statement (i) of Lemma \ref{lemma:unique-subFSR-01-pre},
$\CycleStr{g}\subset\CycleStr{p_1}$.
Denote
\begin{align*}
V_0=&\set{\state{v}\in\BnrSet^{2\len}:
\state{v}\in\cycsecset{2\len}{\cycle{c}},\cycle{c}\in\CycleStr{g}\cap\CycleStr{p_0}};\\
V_1=&\set{\state{v}\in\BnrSet^{2\len}:
\state{v}\in\cycsecset{2\len}{\cycle{c}},\cycle{c}\in\CycleStr{g}\cap\veccomplement{\CycleStr{p_0}}}.
\end{align*}
Since $b=1$ in Eq.(\ref{eqn:order-cnt-dec}),
by Lemma \ref{lemma:cycle-structure-lfsr-p0x},
$\CycleStr{g}$ consists of $[0]$
and $({2^{2\len}-1})/({3\len})$ $3\len$-cycles.
Moreover, by Statement (ii) of Lemma \ref{lemma:unique-subFSR-01-pre},
we have
\begin{equation*}
\cset{\CycleStr{g}\cap\veccomplement{\CycleStr{p_0}}}\ge1,
\end{equation*}
implying $V_1\neq\emptyset$.
Besides, as the  states of the $2\len$-stage FSR $g$,
$V_0\cup V_1=\BnrSet^{2\len}$
and $V_0\cap V_1=\emptyset$.
For $V\subset\BnrSet^{2\len}$,
denote $L(V)=\set{L(\state{v}):\state{v}\in V}$.
On one hand, 
by Lemma \ref{lemma:cycle-state-cycle}, $L(V_0)=V_0$.
Because $L$ is bijective on $\BnrSet^{2\len}$,
we have $L(V_1)=V_1$.
Denote $\state{w}_0=(0,\dots,0,1)\in\BnrSet^{2\len}$.
On the other hand,
by Statements (iii) and (v) of Lemma \ref{lemma:cycle-2-family},
we have 
$L(\state{v})\oplus \state{w}_0\in V_1$
for any $\state{v}\in V_1$.
Thus, both $\state{v}\mapsto L(\state{v})$ and
$\state{v}\mapsto L(\state{v})\oplus \state{w}_0$ are closed on $V_1$.
Since the linear transformation  $L$
has its irreducible minimal polynomial $p_0$ of degree $2\len$,
$L^i(\state{w}_0)$, $i=0,\dots,2\len-1$, is a basis of
the linear space $\BnrSet^{2\len}$.
Then for any $\state{v}_0\in V_1$,
there exist $b_i\in\BnrSet$, $1\leq i\leq 3\len$,
satisfying $\state{v}_0=\bigoplus_{i=1}^{3\len}b_i\cdot L^{3\len-i}(\state{w}_0)$.
Let $\state{v}_i=L(\state{v}_{i-1})\oplus (b_i\cdot\state{w}_0)$, $1\leq i\leq 3\len$.
Then $\state{v}_i\in V_1$,  $1\leq i\leq 3\len$.
However, by Lemmas \ref{lemma:cycle-state-cycle} and \ref{lemma:cycle-structure-lfsr-p0x},
$L^{3\len}$ is an identity map. Hence, $\state{v}_{3\len}=L^{3\len}(\state{v}_0)\oplus
\left(\bigoplus_{i=1}^{3\len}b_i\cdot L^{3\len-i}(\state{w}_0)\right)=\vecnull{2\len}\in V_0$,
yielding $\vecnull{2\len}\in V_0\cap V_1=\emptyset$,
which is ridiculous. Therefore, Case (i) does not occur.

Case(ii).
 $g$ is of stage $1$. Since $[0]\in\CycleStr{g}$, we have
 $\CycleStr{g}=\set{[0],[1]}$, i.e., $g$ is the the LFSR $x_1\oplus x_0$.
\fp

\begin{lemma}\label{lemma:unique-factor-01}
If $f_2$ is satisfiable, then for any 
FSR $h$,
$f\neq h\cascade(x_1\oplus x_0)$.
\end{lemma}
\pf
Assume $f= h\cascade(x_1\oplus x_0)$.
Then $h$ is a $2\len$-stage FSR
and $h(x_0,x_1,\dots,x_{2\len})=x_{2\len}\oplus h_1(x_0,x_1,\dots,x_{2\len-1})$,
where $h_1$ is a $2\len$-input Boolean logic.
{
By Statement (ii) of Lemma \ref{lemma:unique-subFSR-01-pre},
if $f_2$ is satisfiable, then
there exists $\state{v}_0\in 
\BnrSet^{2\len+1}
$
satisfying $f_3(\lbit{2\len}{\state{v}_0})=1$
and $\cycchar{\state{v}_0}=0$.}
Let $f_1$ denote the feedback logic of $f$ and
$\state{v}_0=(a_0,a_1,\dots,a_{2\len})$.
Then $f_1(\state{v}_0)=a_1\oplus a_{\len+1}\oplus\cycchar{\state{v}_0}\oplus
f_3(\lbit{2\len}{\state{v}_0})
=a_1\oplus a_{\len+1}\oplus1$.
Thus,
$f(\state{v}_0\parallel f_1(\state{v}_0))=
h\left(\cycctrl{\state{v_0}}\parallel
(a_{2\len}\oplus a_1\oplus a_{\len+1}\oplus1)\right)=0$,
yielding
\begin{equation}\label{eqn:unique-factor-01-1}
h_1(\cycctrl{\state{v_0}})=a_{2\len}\oplus a_1\oplus a_{\len+1}\oplus1.
\end{equation} 

Let $\state{u}_0=\vecconjugate{\veccomplement{\state{v}_0}}$.
By Statements (i) and (ii) of Lemma \ref{lemma:conjugate-property},
$\cycchar{\state{u}_0}=0$
and $\cycctrl{\state{u}_0}=\vecconjugate{\cycctrl{\state{v}_0}}$.
If
\begin{equation*}
\cycctrl{\state{u}_0}
\neq\min\set{L^i(\cycctrl{\state{u}_0}):
1\leq i\leq 3\len},
\end{equation*}
then
$f_3(\lbit{2\len}{\veccomplement{\state{v}_0}})=f_3(\lbit{2\len}{\state{u}_0})=0$.
Otherwise, assume
$\cycctrl{\state{u}_0}=\min\set{L^i(\cycctrl{\state{u}_0}):
1\leq i\leq 3\len}$.
Since $f_3(\lbit{2\len}{\state{v}_0})=1$, we get
\begin{equation*}
\cycctrl{\state{v}_0}=
\min\set{L^i(\cycctrl{\state{v}_0}):1\leq i\leq 3\len}.
\end{equation*}
As $\cycctrl{\state{u}_0}=\vecconjugate{\cycctrl{\state{v}_0}}$,
by Lemmas \ref{lemma:conjugate-nominkiss}
and \ref{lemma:cycle-structure-lfsr-p0}, we have
$\set{\cycctrl{\state{v}_0},\cycctrl{\state{u}_0}}=\set{\vecnull{2\len},\veconenull{2\len}}$.
Considering $\cycchar{\state{v}_0}=\cycchar{\state{u}_0}=0$,
we have
\begin{equation*}
\set{{\state{v}_0},{\state{u}_0}}=\set{\vecnull{2\len+1},
\veccomplement{\veconenull{2\len+1}}}.
\end{equation*}
Because $f_3(\lbit{2\len}{\state{v}_0})=1$
while $f_3(\vecnull{2\len})=0$,
we have $\state{u}_0=\vecnull{2\len+1}$,
yielding $f_3(\lbit{2\len}{\veccomplement{\state{v}_0}})=
f_3(\lbit{2\len}{\state{u}_0})=0$.

We have proved $f_3(\lbit{2\len}{\veccomplement{\state{v}_0}})=0$.
Then $F(\veccomplement{\state{v}_0})=L_1(\veccomplement{\state{v}_0})$,
where $F$ is the state transformation of $f$.
Using $\cycchar{{\state{v}_0}}=0$ and Statements (i)-(ii) of Lemma \ref{lemma:conjugate-property},
we get
\begin{equation*}
f(\veccomplement{\state{v}_0}\parallel (a_1\oplus a_{\len+1}\oplus\cycchar{\veccomplement{\state{v}_0}}\oplus f_3(\lbit{2\len}{\veccomplement{\state{v}_0}})))=
h(\cycctrl{\state{v}_0}
\parallel(a_{2\len}\oplus a_1\oplus a_{\len+1}))=0,
\end{equation*}
implying
\begin{equation}\label{eqn:unique-factor-01-2}
h_1(\cycctrl{\state{v}_0})
=a_{2\len}\oplus a_1\oplus a_{\len+1}
.
\end{equation}

Our assumption $f= h\cascade(x_1\oplus x_0)$
leads to contradictory Eqs. (\ref{eqn:unique-factor-01-1}) and (\ref{eqn:unique-factor-01-2}).
The proof is completed.
\fp

\begin{lemma}\label{lemma:decomposable-reducible}\cite{GD70}
Let ${h}$ be an $m$-stage
decomposable FSR satisfying $h(\vecnull{m+1})=0$.
Then there exist two FSRs $h_1$ and $h_2$
such that $h=h_1\cascade h_2$, where $h_2$ is
a $k$-stage FSR for some $1\le k<m$
and $[0]\in\CycleStr{h_2}$.
Particularly, $h_2$ is a subFSR of $h$ and $h$ is reducible.
\end{lemma}
\pf Since $h$ is decomposable,
we assume $h=h_1'\cascade h_2'$,
where $h_2'$ is a $k$-stage FSR, $1\leq k<m$.
If $h_2'(\vecnull{k+1})=0$,
let $h_1=h_1'$ and $h_2=h_2'$.
Assume $h_2'(\vecnull{k+1})=1$.
Let $h_2=h_2'\oplus1$ and
$h_1(x_0,x_1,\dots,x_{m-k})=h_1'(x_0\oplus1,x_1\oplus1,\dots,x_{m-k}\oplus1)$.
Then $h=h_1'\cascade h_2'=h_1\cascade h_2$
and $h_2(\vecnull{k+1})=h_2'(\vecnull{k+1})\oplus1=0$.
Besides, $h_2(\vecnull{k+1})=0$ is equivalent to $[0]\in\CycleStr{h_2}$.

Because
$h_1(\vecnull{m-k+1})=h_1(h_2(\vecnull{k+1}),h_2(\vecnull{k+1}),\dots,h_2(\vecnull{k+1}))=
h(\vecnull{m+1})=0$,
we have $\SeqSet{h_2}\subset\SeqSet{h_1;h_2}=\SeqSet{h}$,
where $\SeqSet{h_1;h_2}$ is the set of sequences generated by
the cascade connection of $h_1$ into $h_2$.
Therefore, $h_2$ is a subFSR of $h$ and $h$ is reducible.
\fp
The idea of Lemma \ref{lemma:decomposable-reducible}
was given by \cite{GD70} and here we reinterpret it for readability.

\begin{lemma}\label{lemma:reduction-indecomposable}
The FSR $f$ is indecomposable
if and only if the Boolean circuit $f_0$ is satisfiable.
\end{lemma}
\pf Consider two cases below.

Case (i):  $f_0$ is satisfiable.
 By Lemma \ref{lemma:f2-f0-satisfiability},
 $f_2$ is satisfiable.
Assume  $f$ to be decomposable.
Since $f_2(\vecnull{r})=0$,
by Algorithm \ref{alg:fsr-dec}, we have
$f_3(\vecnull{2\len})=0$ and 
$f(\vecnull{2\len+1})=0$, implying $[0]\in\CycleStr{f}$.
By Lemma \ref{lemma:decomposable-reducible},
there exist FSRs $h$ and $g$
such that  $f=h\cascade g$,
where $g$ is a subFSR of $f$ satisfying $[0]\in\CycleStr{g}$.
By Lemma \ref{lemma:unique-subFSR-01},
$g$ is the LFSR $x_1\oplus x_0$.
However,
by Lemma \ref{lemma:unique-factor-01},
$f\neq h\cascade(x_1\oplus x_0)$.
Hence, the assumption is absurd
and $f$ is indecomposable.

Case (ii): $f_0$ is unsatisfiable.
By Lemma \ref{lemma:f2-f0-satisfiability},
$f_2$ is unsatisfiable.
By Algorithm \ref{alg:fsr-dec}, $f_3(\state{x})=0$ for any $\state{x}\in\BnrSet^{2\len}$.
Then $f$ is exactly the LFSR $p_1$ and
$f(x_0,x_1,\dots,x_{2\len})=(x_{2\len}\oplus x_{\len}\oplus x_0)\cascade(x_1\oplus x_0)$.
So, $f$ is decomposable.
\fp

\begin{center}
\begin{minipage}[h]{0.7\textwidth}
\textbf{PROBLEM}: FSR INDECOMPOSABILITY

{INSTANCE}: An FSR $f$ with its feedback logic $f_1$ as a Boolean circuit of size $\sizeof{f_1}$.

{QUESTION}: Is $f$ indecomposable?
\end{minipage}
\end{center}

By Lemmas \ref{lemma:CKT-SAT-NP-C},
\ref{lemma:alg-dec-poly-time} and \ref{lemma:reduction-indecomposable},
Algorithm \ref{alg:fsr-dec} is a polynomial-time Karp reduction
from CIRCUIT SATISFIABILITY
to 
FSR INDECOMPOSABILITY.
Therefore, we conclude that
\begin{theorem}\label{thm:indecomposability-np-hard}
The 
FSR INDECOMPOSABILITY
problem is \NP-hard.
\end{theorem}

\section{Conclusion}\label{sect:conclusion}
Deciding irreducibility/indecomposability
of FSRs is meaningful for sophisticated circuit implementation
and security analysis of stream ciphers.
Here we have proved both the decision problems are \NP-hard.
Assuming \textbf{P}$\neq$\NP, where \textbf{P} is the class of decision problems computed by polynomial-time deterministic Turing machines,  it is intractable to
find a polynomial-time computable 
algorithm 
for either problem.

Furthermore, it is still of theoretical interests to determine the computational complexity of
search versions of FSR reducibility/decomposability, i.e.,
to find a subFSR/factor of a given FSR,
where $g$ and $h$ are called factors of $f$ if $f=h\cascade g$.
Besides, provided that the input Boolean circuit is satisfiable,
Algorithm \ref{alg:fsr-red}(resp. Algorithm \ref{alg:fsr-dec})
constructs an irreducible(resp. indecomposable) FSR.
Since it is easy to efficiently find satisfiable Boolean circuits,
it remains a question 
whether
Algorithm \ref{alg:fsr-red}(resp. Algorithm \ref{alg:fsr-dec})
can be modified to construct a family of irreducible(resp. indecomposable)
FSRs with desirable properties in practice.

\section{Appendices}

\subsection{Appendix: the proof of Statement (i) of Lemma \ref{lemma:cycle-join}}\label{appnd:proof-cycle-join}
\pf
Let  $F$
denote the state transformation of the FSR $f$.

By Lemma \ref{lemma:cycle-state-cycle},
it is sufficient to prove the following claim.

\emph{Claim:}
For any $\state{u},\state{v}\in \BnrSet^m$,
there exists $i\geq0$ satisfying $F^i(\state{u})=\state{v}$
if and only if $\state{u},\state{v}\in\bigcup_{\cycle{c}\in\CycleSet{C}}\cycsecset{m}{\cycle{c}}$,
where $\CycleSet{C}$ is a weakly connected component of $D_g^f$.

We prove this claim by induction on the number of arcs in $D_g^f$.

If $D_g^f$ has no arc, then by Eq.(\ref{eqn:cycle-join-lambda-cnd}),
$f_3(\lbit{m-1}{\state{v}})=0$ for any $\state{v}\in\BnrSet^m$.
Thus, $\CycleStr{g}=\CycleStr{f}$ and the claim holds.

Now suppose that $D_g^f$ has at least one arc.

Because $D_g^f$ is acyclic, there exists a source
$\cycle{c}_0\in\CycleStr{g}$ with positive outdegree.
Denote $V=\set{\state{v}\in\cycsecset{m}{\cycle{c}_0}:f_3(\lbit{m-1}{\state{v}})=1,
\cycextr{{\state{v}}}=1}$.
By Eq.(\ref{eqn:cycle-join-lambda-cnd}),
$\cset{V}=1$ and there is a unique arc leaving $\cycle{c}_0$.
Denote $V=\set{\state{v}_0}$.
Let $\cycle{c}_1
$ denote the unique successor of $\cycle{c}_0$,
and let
$\CycleSet{C}$ denote the weakly connected component
containing $\cycle{c}_0$.
We have $\cycle{c}_1\neq\cycle{c}_0$ because $D_g^f$ is acyclic.

Let ${\state{v}_0}=(v_0,v_1,\dots,v_{m-1})$ and
\begin{align*}
f_3'(x_1,\dots,x_m)=&f_3(x_1,\dots,x_m)\oplus\prod_{i=1}^{m-1}(x_i\oplus v_i\oplus1);\\
f'(x_0,x_1,\dots,x_m)=&g(x_0,x_1,\dots,x_m)\oplus f_3'(x_1,\dots,x_m).
\end{align*}
Define a directed graph $D_g^{f'}$ with the set of vertices $\CycleStr{g}$
such that an arc is incident from $\cycle{a}$ to $\cycle{b}$ if and only if
\begin{equation*}
\set{\state{v}\in\cycsecset{m}{\cycle{a}}:
f_3'(\lbit{m-1}{\state{v}})=1,
\cycextr{\state{v}}=1,
\vecconjugate{\state{v}}\in \cycsecset{m}{\cycle{b}}}\neq\emptyset.
\end{equation*}
See that $f_3'$ differs from $f_3$  only at $(v_1,\dots,v_{m-1})$
with $f_3'(v_1,\dots,v_{m-1})=0$.
Then $D_g^{f'}$ is obtained by removing the arc leaving
$\cycle{c}_0$ in $D_g^f$.
Besides,
Eq.(\ref{eqn:cycle-join-lambda-cnd}) also holds for $f_3'$.

Denote $F'$ as the state transformation of $f'$.
The cycle joining method gives
\begin{equation} \label{eqn:cycle-join-interchange-GG}
F'(\state{v})=
\left\{
\begin{aligned}
 F(\vecconjugate{\state{v}}),&\text{ if } {\state{v}}\in\set{\state{v}_0,\vecconjugate{\state{v}_0}}; \\
  F(\state{v}),& \text{ otherwise.}
 \end{aligned}
\right.
\end{equation}

By induction, the claim above is assumed to hold for $f'$.
We only have to consider states in $\bigcup_{\cycle{c}\in\CycleSet{C}}\cycsecset{m}{\cycle{c}}$.
In $D_g^{f'}$, $\CycleSet{C}\setminus\set{\cycle{c}_0}$
and $\set{\cycle{c}_0}$
are weakly connected components.
Denoting $p=\lengthof{\cycle{c}_0}$
and $q=\sum_{\cycle{c}_0\neq\cycle{c}\in\CycleSet{C}}\lengthof{\cycle{c}}$,
and using Lemma \ref{lemma:cycle-state-cycle},
we have
\begin{equation}\label{eqn:cycle-join-subcycles}
\left\{
\begin{aligned}
&\set{F'^i(F({\state{v}_0})):0\leq i< q}=
\bigcup_{\cycle{c}_0\neq\cycle{c}\in\CycleSet{C}}\cycsecset{m}{\cycle{c}};\\
&\set{F'^i(F(\vecconjugate{\state{v}_0})):0\leq i< p}=\cycsecset{m}{\cycle{c}_0};\\
&F'^{q-1}(F({\state{v}_0}))=\vecconjugate{\state{v}_0};\\
&F'^{p-1}(F(\vecconjugate{\state{v}_0}))={\state{v}_0}.
\end{aligned}
\right.
\end{equation}
By Eqs. (\ref{eqn:cycle-join-interchange-GG}) and (\ref{eqn:cycle-join-subcycles}),
$F^{p+q}({\state{v}_0})=\state{v}_0$ and
\begin{equation*}
    \set{F^i({\state{v}_0}):0\leq i< p+q}=
    \bigcup_{\cycle{c}\in\CycleSet{C}}\cycsecset{m}{\cycle{c}}.
\end{equation*}
Thus, the claim also holds for $f$.

The proof of this claim is complete by induction.
\fp

\subsection{Appendix: The operation $\min$}\label{appnd:min}
The operation \textquotedblleft$\min$\textquotedblright\,
outputs the minimum of two integers.

\begin{figure}[htb]
\begin{center}
\setlength{\unitlength}{1.7mm}
\begin{picture}(43.5,24) 
\put(25.75,8){\usebox{\myMIN}}
\end{picture}
\end{center}
\caption{A recursive construction of the Boolean circuit $\min_m$}\label{fig:max}
\end{figure}
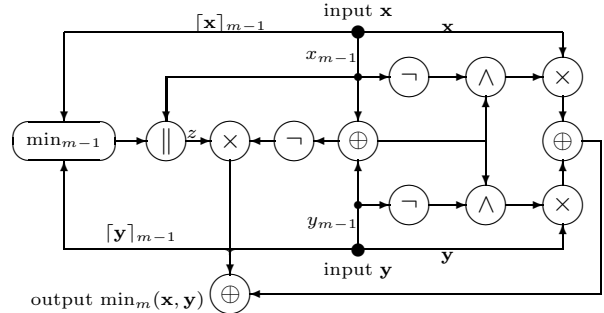

Let $\min_{m}$ denote the operation computing
the minimum of two $m$-bit nonnegative integers.
Recall that a vector $\state{v}=(v_0,v_1,\dots,v_{m-1})$
is identified as the integer $\sum_{i=0}^{m-1}v_i2^i$.
For $m=1$, we have $\min_1(x_{0},y_{0})=x_0\myAND y_0$.
For $m\geq 2$, $\state{x}=(x_0,x_1,\dots,x_{m-1})$ and
$\state{y}=(y_0,y_1,\dots,y_{m-1})$, we have
\begin{align*}
\min{_m}(\state{x},\state{y})= &
(x_{m-1}\oplus y_{m-1}\oplus1)
\times(\min{}_{m-1}(\hbit{m-1}{\state{x}},\hbit{m-1}{\state{y}})
\parallel x_{m-1})\\
&\oplus (((x_{m-1}\oplus y_{m-1}) \myAND (x_{m-1}\oplus1))\times \state{x})\\
&\oplus (((x_{m-1}\oplus y_{m-1}) \myAND (y_{m-1}\oplus1))\times \state{y}),
\end{align*}
and thereby give a recursive description of 
$\min_m$ in Figure \ref{fig:max},
where $z=\min{_{m-1}}(\hbit{m-1}{\state{x}},\hbit{m-1}{\state{y}})
\parallel x_{m-1}$.
Here the multiplying operation \textquotedblleft$\times$\textquotedblright\, has a one-bit input $a$
and an $m$-bit input $\state{w}=(w_0,w_1,\dots,w_{m-1})$,
and outputs $(a\myAND w_0,a\myAND w_1,\dots,a\myAND w_{m-1})$.
Thus, the multiplying operation \textquotedblleft$\times$\textquotedblright\,
costs $m$ gates.
By Figure \ref{fig:max}, we have
$\sizeof{\min_m}=12+13m+\sizeof{\min_{m-1}}$ for any $m\geq2$,
and hence
$\sizeof{\min_{m}} = (13m^2+37m-44)/2$.

\end{document}